%% file: vep-article.tex
\documentclass[12pt,draft]{article}

\usepackage{amsmath}
\usepackage{graphicx}
\usepackage{feynmf}
\usepackage{color}

\newcommand\mysection{\setcounter{equation}{0}\section}

\newcounter{hran}

\newcommand{\cals}{\mbox{${\cal S}$}}

\newcommand{\deltal}{\Delta_{\alpha}}
\newcommand{\bbar}{\bar{b}}
\newcommand{\detg}{\det(G)}
\newcommand{\dets}{\det(\cals)}

\begin{document}

\begin{fmffile}{samplepics}
\setlength{\unitlength}{1mm}

\begin{titlepage}

\vspace{1.cm}

\begin{center}

{\large \bf Stable \\
One-Dimensional Integral Representations \\
of One-Loop 
$N$-Point Functions \\
\vspace{0.1cm}
in the General Massive Case\\[1cm]
I - Three Point Functions}\\[2cm]

{\large  J.~Ph.~Guillet$^{a}$, E.~Pilon$^{a}$, 
M.~Rodgers$^{b}$ and M. S. Zidi$^{a, c}$ } \\[.5cm]

\normalsize
{$^{a}$ LAPTh, Universit\'e de Savoie, CNRS, B.P. 110, Annecy-Le-Vieux, 
F-74941, France}\\
{$^{b}$ IPPP, Department of Physics, Durham University, Durham, DH1 3LE, United Kingdom}\\
{$^{c}$ LPTh, Universit\'e de Jijel, B.P. 98 Ouled-Aissa, 18000 Jijel, Alg\'erie}\\
      
\end{center}

\vspace{2cm}

%

\begin{abstract} 
\noindent
In this article we provide representations for the one-loop three point 
functions in 4 and 6 dimensions in the general case with complex masses. 
The latter are part of the {\tt GOLEM} library used for the computation of 
one-loop multileg amplitudes. 
These representations are one-dimensional integrals designed to be free of 
instabilites induced by inverse powers of Gram determinants, therefore
suitable for stable numerical implementations.    
\end{abstract}

\vspace{1cm}

\begin{flushright}
LAPTH-060/13\\
IPPP/13/69 DCPT/13/138
\end{flushright}

\vspace{2cm}

\end{titlepage}

\pagestyle{plain} 

\input{vep_intro}
\newpage
\input{vep_3-point}

\input{concl}

\appendix

\input{appendix-magicidientities}

\input{appendix-detg-zero-kinematics}

\input{appendix-spectral-decomp-S}

\input{appendix-calc-bj-B}

\input{appendix-zeromode}

\input{vep_appendix_b}

\vspace{0.5cm}

\noindent

\input{bibliography}
\end{fmffile}

\end{document}

%% file: vep_intro.tex
\mysection{Introduction}\label{sect1}

The Golem project \cite{golem} initially aimed at automatically computing one loop
corrections to QCD processes using Feynman diagrams techniques whereby 1) each 
diagram was written as form factors times Lorentz structures 2) each form factor 
was decomposed on a particular redundant set of basic integrals. Indeed when the form 
factors are reduced down to a basis of scalar integrals only, negative powers of 
Gram determinants, generically noted $\det(G)$ below, show up in separate 
coefficients of the decomposition. These $\det(G)$, albeit spurious, are sources
of troublesome numerical instabilities whenever they become small. The set of
basic integrals used in the Golem approach is such that all coefficients of the
decomposition of any form factor on this set are free of negative powers of
$\det(G)$. Let aside trivial one- and  two-point functions, the Golem library of
basic functions is instead made of a redundant set involving the functions
$I_3^n(j_1,\cdots,j_3)$, $I_3^{n+2}(j_1)$,  $I_4^{n+2}(j_1,\cdots, j_3)$ and
$I_4^{n+4}(j_1)$. Here the lower indices indicate the number of external legs,
the upper indices stand for the dimension  of space-time, and the arguments 
$j_1, \cdots,j_i$ labels $i$ Feynman parameters in the numerator of the 
corresponding integrand. The strategy is the following. In the phase space 
regions where $\det(G)$ are not troublesome, the extra elements of the Golem 
set are decomposed on a scalar basis and computed ``analytically'' in terms of 
logarithms and dilogarithms. In the phase space region where $\det(G)$ vanishes 
these extra Golem elements are instead used as irreducible building
blocks explicitly free of Gram determinant and provided as one-dimensional 
integral representations computed ``numerically''.

\vspace{0.3cm}

\noindent
Much faster and more efficient methods than those relying on Feynman diagrams 
techniques  have been developed, e.g. based on unitarity cuts of transition 
amplitudes and not individual Feynman diagrams, and/or 
processing the decompositions at the level of the integrands \cite{bdk,britt,opp}. 
Yet these methods still amount to a decomposition onto a set of 
basic integrals. In this respect the stand-alone relevance of the Golem library 
of basic functions, initially developed as a part of the Golem approach, remains. 
Furthermore the decompositions obtained by these new methods project onto a 
basis of scalar integrals and thus are still submitted to numerical 
instabilities caused by $\det(G)$. The issue of numerical instability is then 
addressed in various ways ranging from smoothing numerical interpolations over
the regions of instabilities \cite{papa1} to more involved rescue solutions \cite{ellis,denn}. In \cite{gosam}
the solution adopted is to provide a rescue alternative relying on the Golem
decomposition to compute the amplitude in the troublesome kinematic
configurations. The Golem library \cite{golem95}, initially designed for QCD, did not include
basic functions with internal masses yet provided a convenient way of handling 
infrared and collinear singularities inherent in the massless case. Its
completion with the cases involving internal masses, possibly complex, extends
its range of use \cite{golem95c}. This completion shall supply the functions
$I_3^n(j_1,\cdots,j_3)$, $I_3^{n+2}(j_1)$, $I_4^{n+2}(j_1,\cdots, j_3)$ and 
$I_4^{n+4}(j_1)$ in the massive cases  in a numerically stable with respect to
$\det(G)$ issues.   

\vspace{0.3cm}

\noindent
To handle $\det(G)$ issues, we advocate the use of one-dimensional 
integral representations rather than relying on Taylor expansions in powers of 
$\det(G)$. The latter may be thought a priori better both in terms of CPU time
and accuracy, however the order up to which the expansion shall be
pushed may happen to be rather large. Furthermore, unless a fixed large 
number of terms, hopefully large enough in all practical cases, be computed, it
is not easy to assess a priori the optimal order required to reach a given
accuracy. Actually this assessment would demand a quantitative estimate of the
remainder as a function of the order of truncation, which, as with the Taylor
expansion with Laplace remainder, namely requires the computation of an
integral! Originally, we proposed the antipodal option of computing numerically 
the two- or three-dimensional Feynman integral defining respectively the
three- and four point functions, more precisely hypercontour deformations 
thereof \cite{golem} that would be numerically more stable. Yet the computation
of these multiple integrals was both slow and not very precise. It is far more
efficient both in terms of CPU time and accuracy to evaluate a one-dimensional
integral representation, insofar as one is able to find such a representation.
In the case without internal masses, we indeed found such a representation.

\vspace{0.3cm}

\noindent
The issue which we address here is the extension of this approach of 
one-dimensional integral representations for our set of basic integrals in the
most general case, i.e. with internal complex masses. In this article
we treat the case of the three point function.
The case of four point functions is more involved therefore it will be 
elaborated separately in a companion article. We follow the 
approach developed by t'Hooft and Veltman in ref. \cite{hv}. 
In a subsequent third article, we will present 
an alternative approach providing integral representations for both three and 
four point functions equivalent to the one presented here yet with a number of 
new features and advantages. 
The present article is organized as follows. Section \ref{sec_3point}
sketches the derivation of the three point function leading to our integral 
representation. Section \ref{sect3} treats the case when $\det (G)$ vanishes
whereas the determinant of the kinematic matrix ${\cal S}$ remains non 
vanishing. Section \ref{sect4} elaborates on the more tricky case when both 
$\det (G)$ and the $\det ({\cal S})$ vanish. The main body of the text presents
the general arguments whereas the various technical details supporting the
latter are gathered in appendices, to make the reading of this article more 
fluent. 

%% file: vep_3-point.tex
\mysection{Outline of the derivation}\label{sec_3point}

\noindent
A generic three point function can be 
represented by the diagram of Fig. \ref{fig1}:


\begin{figure}[h]
\centering
\parbox[c][43mm][t]{80mm}{\begin{fmfgraph*}(60,80)
  \fmfleftn{i}{1} \fmfrightn{o}{1} \fmftopn{t}{1}
  \fmf{fermion,label=$p_1$}{t1,v1}
  \fmf{fermion,label=$p_2$}{i1,v2}
  \fmf{fermion,label=$p_3$}{o1,v3}
  \fmf{fermion,tension=0.5,label=$q_1$}{v1,v2}
  \fmf{fermion,tension=0.5,label=$q_2$}{v2,v3}
  \fmf{fermion,tension=0.5,label=$q_3$}{v3,v1}
\end{fmfgraph*}}
\caption{The triangle picturing the one-loop three point function.}
\label{fig1} 
\end{figure}
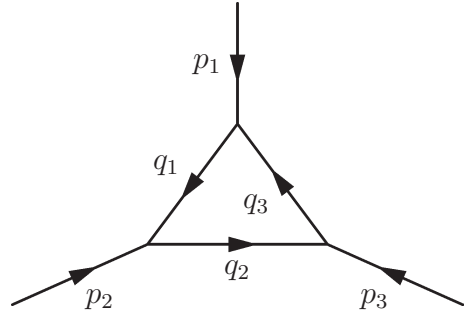

Each internal line with momentum $q_i$ stands for the propagator of a particle
of mass $m_i$. We define the kinematic matrix $\cals$, which 
encodes all the information on the kinematics associated to this diagram by:
\begin{equation}
\cals_{i \, j} = (q_i-q_j)^2 - m_i^2 - m_j^2
\label{eqDEFCALS}
\end{equation}
The squares of differences of two internal momenta can be written in terms of
the internal masses $m_i$ and the external invariants $s_i = p_i^2$ so that 
$\cals$ reads:
\begin{equation}\label{eqcals}
\cals 
= 
\left[ 
 \begin{array}{ccc}
 -2 \, m_1^2 & s_2 - m_1^2 - m_2^2 & s_1 - m_1^2 - m_3^2 \\
 s_2 - m_1^2 - m_2^2 & - 2 \, m_2^2 & s_3 - m_2^2 - m_3^2 \\
 s_1 - m_1^2 - m_3^2 & s_3 - m_2^2 - m_3^2 & - 2 \, m_3^2
 \end{array}
\right]
\end{equation}

In this section, we will sketch the computation of $I^4_3$ and $I^6_3$
using the method of ref. \cite{hv}. 
These two integrals are defined\footnote{The Feynman contour prescription in
the propagators is noted $i \lambda$ in order to avoid any confusion with the
parameter $\epsilon = (4-n)/2$ involved in dimensional regularization.} by:
\begin{eqnarray}
I^4_3 
& = & -
\int_0^1 \, \prod_{i=1}^3 \, d z_i \, 
\delta \left( 1- \sum_{i=1}^3 \, z_i \right) 
\left(  -\frac{1}{2} \, z^{\;T} \, \cals \, z - i \, \lambda \right)^{-1}
\label{eqSTARTINGPOINT}
\end{eqnarray}
\begin{eqnarray}
I^{n+2}_3 
& = & 
- \, \frac{\Gamma(1+\epsilon)}{\epsilon} 
\int_0^1 \, \prod_{i=1}^3 \, d z_i \, 
\delta \left( 1- \sum_{i=1}^3 \, z_i \right) 
\left(  
- \,\frac{1}{2} \, z^{\;T} \, \cals \, z - i \, \lambda 
\right)^{-\epsilon} 
\nonumber \\
& = & 
- \, \frac{\Gamma(1+\epsilon)}{\epsilon} \, 
\int_0^1 \, \prod_{i=1}^3 \, d z_i \, 
\delta \left( 1- \sum_{i=1}^3 \, z_i \right) 
\left[ 
 1 - \epsilon \, 
 \ln \left(  -\frac{1}{2} \, z^{\;T} \, \cals \, z - i \, \lambda \right) 
\right] 
\nonumber \\
& = & 
I_3^{div} + I^6_3
\label{eqSTARTINGPOINT6}
\end{eqnarray}
where $I_{3}^{div}$ isolates the $\overline{MS}$ ultra violet pole in
$\epsilon$, and $I^6_3$ is the finite part which we will focus on. 
We may single out any index $a$ in  $S =\{1,2,3\}$ and write 
\begin{equation}\label{singleout}
z_a = 1 - \sum_{i \ne a} z_i
\end{equation}
The quadratic form $z^{\;T} \, \cals \, z$ becomes:
\begin{eqnarray}
z^{\;T} \, \cals \, z 
& = & 
- \sum_{i,j \ne a} G_{i\,j}^{(a)} \, z_i \, z_j + 
2 \, \sum_{j \ne a} V_j^{(a)} \, z_j + \cals_{a\,a}
\label{eqVECZT}
\end{eqnarray}
with 
\begin{eqnarray}
G_{i\,j}^{(a)} 
& = & 
- (\cals_{i\,j}-\cals_{a\,j}-\cals_{i\,a}+\cals_{a\,a}), \;\; i,j \neq a
\label{eqGRAMMAT} \\
V_j^{(a)} 
& = & 
\cals_{a\,j} - 
\cals_{a\,a} \;\;\;\;\;\;\;\;\;\;\;\;\;\;\;\;\;\;\;\;\;\;\;\;\;\;\;\;\;\; 
j \neq a
\label{eqVJA}
\end{eqnarray}
The matrix $G^{(a)}$ is the $2 \times 2$ Gram matrix built from the 
four-vectors $\Delta_{i \, a} = q_i-q_a$: 
$G^{(a)}_{ij} = 2(\Delta_{i \, a}.\Delta_{j \, a})$. 
Its determinant does not depend on the choice of $a$, and it is also the
determinant of the similar Gram matrix built with any subset of two external 
momenta. We note it simply $\detg$ without 
referring to $a$ and unambiguously call it {\em the} Gram determinant associated 
with the kinematic matrix $\cals$.
Specifying for example $a=3$, $I^4_3$ reads:
\begin{eqnarray}
I^4_3 
& = & 
- \int_0^1 d z_1 \, \int^{1-z_1}_0 d z_2 \,
\left[ 
 \frac{1}{2} \, 
 \sum_{i,j = 1}^{2} G^{(3)}_{ij} \, z_i \, z_j \, 
 - \sum_{j = 1}^{2} V_{j}^{(3)} \, z_j \, - \, 
 \frac{1}{2} \cals_{a\,a} \, - i \, \lambda 
\right]^{-1}
\label{eqI30}
\end{eqnarray}
The $2 \times 2$ Gram matrix $G^{(3)}$ and the column two-vector $V^{(3)}$ are 
explicitly given by:
\begin{eqnarray}
G^{(3)} 
& = & 
\left[ 
 \begin{array}{cc}
 2 \, s_1 & s_3 - s_2 + s_1 \\
 s_3 - s_2 + s_1 & 2 \, s_3 
 \end{array}
\right] 
\label{eqDEFG3} \\
V^{(3)} 
& = & 
\left[ 
 \begin{array}{c}
 s_1 - m_1^2+m_3^2 \\
 s_3 - m_2^2+m_3^2
 \end{array}
\right] 
\label{eqDEFV3}
\end{eqnarray}
We then define
\begin{eqnarray*}
z_1 & = & 1-x \\
z_2 & = & y 
\end{eqnarray*}
and we get\footnote{The argument of the logarithm appearing in
eq. (\ref{eqI361}) shall be understood to contain an implicit arbitrary factor 
$1/M^2$ with dimension -2 in order to make the argument of this logarithm 
dimensionless. This arbitrary $M^2$ dependence is cancelled by the 
corresponding one in the $\ln (M^2/\mu^2)$ involved in the term $I_3^{div}$, 
where $\mu^2$ is the dimension two parameter introduced by the dimensional 
regularization of the ultra violet divergence subtracted in $I_3^{div}$. 
In practice the kinematic matrix $\cals$ is rescaled from the start by its 
entry of largest absolute value, and so is the Gram matrix $G^{(a)}$, which
thereby become both dimensionless. This amounts to specifying $M^2$ to be this 
normalization parameter.}:
 \begin{eqnarray}
I^4_3
& = & 
-\int_0^1 d x \, \int^{x}_0 d y \,
\left[ 
  a \, x^2 + b \, y^2 + c \, x \, y  + d \, x + e \, y + f - i \, \lambda 
\right]^{-1}
\label{eqI31} \\
I^6_3
& = & 
\int_0^1 d x \, \int^{x}_0 d y \,
\ln 
\left[ 
  a \, x^2 + b \, y^2 + c \, x \, y  + d \, x + e \, y + f - i \, \lambda 
\right]
\label{eqI361}
\end{eqnarray}
with:
\begin{equation}\label{list}
\begin{array}{l}
a = s_1 \\
b = s_3 \\
c = -s_3 + s_2 - s_1 \\
d = m_3^2 - m_1^2 - s_1 \\
e = s_1 - s_2 + m_2^2 - m_3^2 \\
f = m_1^2
\end{array}
\end{equation}
Eq. (\ref{eqI31}) is the starting point of the computation of the 
three point integral in ref. \cite{hv}, c.f. their eq (5.2). 
We keep the same notations as those of ref. \cite{hv} for the different 
quantities, and  we closely follow the strategy of ref. \cite{hv} 
for the first integration. We only sketch these stages. An alternative
strategy may be proposed which leads to the sought integral 
representations in a faster, more straightforward and more transparent way for
three point functions, and which can be elaborated for four point functions as
well thereby providing a number of interesting features. 
This alternative will be presented in a subsequent publication.

\vspace{0.3cm}

\noindent
The integration variable $y$ is first shifted according to 
$y = y^{\prime} +\alpha \, x$, the parameter $\alpha$ is being chosen such that 
\begin{equation}\label{eqTRINOME-ALPHA}
b\, \alpha^2 + c \, \alpha + a = 0
\end{equation} 
in order that the quadratic form of $x,y$ in the integrands of eqs. 
(\ref{eqI31}), (\ref{eqI361}) become linear in $x$. Note that the 
discriminant $\Delta_{\alpha}$ of eq. (\ref{eqTRINOME-ALPHA}) is
minus the Gram determinant $\detg$. 
For all kinematical configurations $p_{1},p_{2},p_{3}=-p_{1}-p_{2}$ involved in 
one-loop calculations of elementary processes of interest  for collider physics,
$\detg$ is non-positive\footnote{As seen by exhaustion, the only 
configurations leading to a positive Gram determinant would require that all
three external four-momenta $p_{1},p_{2},p_{3}=-p_{1}-p_{2}$ of the three point
function be spacelike. At the one-loop order which is our present concern, each
of the three points, through which $p_{1},p_{2}$ and $p_{3}$ respectively flow, 
shall be connected to an independent tree. In order for $p_{1},p_{2}$ and 
$p_{3}$ to be all space-like, each of these trees should involve one leg in the 
initial state: this would correspond neither to a decay nor to a collision of 
two incoming bodies.}. The roots $\alpha_{\pm}$ of
the polynomial  (\ref{eqTRINOME-ALPHA}) are thus real in all relevant cases. We
split the integral over $y^{\prime}$ and reverse the order of integrations:
\begin{eqnarray}
\int_{0}^{1} dx \, \int_{- \alpha}^{1 - \alpha}dy^{\prime} 
& = & 
\int_{0}^{1} dx \, \int_{0}^{(1 - \alpha) x} dy^{\prime} 
\, - \, 
\int_{0}^{1} dx \, \int_{0}^{- \alpha x} dy^{\prime} 
\nonumber\\
& = & 
\int_{0}^{1 - \alpha} dy^{\prime}  \, \int_{y^{\prime}/(1 - \alpha)}^{1} dx 
\, - \, 
\int_{0}^{- \alpha} dy^{\prime}    \, \int_{y^{\prime}/(- \alpha)}^{1} dx
\label{change}
\end{eqnarray}
Since the integrand seen as a function of $x$ and $y^{\prime}$ in eq. 
(\ref{change}) is now linear in $x$ the integration on $x$ is made 
straightforward. 
For $I_{3}^{4}$ eq. (\ref{change}) involves two integrals of the form
\begin{equation}\label{interm1}
\int_{x_{min}(y^{\prime})}^{1} dx \left[ {\cal A} \, x + {\cal B} \right] ^{-1} 
=
\frac{1}{\cal A} 
\ln \left( \frac{{\cal A} + {\cal B}}{{\cal A} \, x_{min}+ {\cal B}} \right)
\end{equation}
where ${\cal A}$ and ${\cal B}$ are functions of $y^{\prime}$ and 
$x_{min}(y^{\prime}) = y^{\prime}/(1-\alpha)$ and 
$x_{min}(y^{\prime}) = y^{\prime}/(-\alpha)$ respectively. As can be traced 
back to eqs. 
(\ref{eqSTARTINGPOINT}), (\ref{eqSTARTINGPOINT6}) the polynomial 
$[a \, x^2 + b \, y^2 + c \, x \, y  + d \, x + e \, y + f - i \, \lambda ]$
in eqs. (\ref{eqI31}), (\ref{eqI361}) has a negative imaginary part, this holds 
true also for complex internal
masses. Therefore the numerator and denominator in the argument of the
logarithm in eq. (\ref{interm1}) both have a negative imaginary part, thus the 
logarithm in eq. (\ref{interm1}) can be harmlessly split in two terms:
\begin{equation}\label{change2}
\ln \left( \frac{{\cal A} + {\cal B}}{{\cal A} \, x_{min}+ {\cal B}} \right)
= 
\ln \left( {\cal A} + {\cal B} \right) \, - \, 
\ln \left( {\cal A} \, x_{min}(y^{\prime})+ {\cal B} \right)
\end{equation}
It is convenient to add and subtract a term $\ln ({\cal C})$ in 
the right hand side (r.h.s.) of eq. (\ref{change2}), and split the latter into 
a sum of two terms
\begin{eqnarray}
\lefteqn{\int_{x_{min}}^{1} dx \left[ {\cal A} \, x + {\cal B} \right] ^{-1}}
\nonumber\\ 
& = &
\frac{1}{\cal A} 
\left[ 
 \ln \left( {\cal A} + {\cal B} \right) -  \ln \left( {\cal C} \right)
\right] 
\, - \, 
\frac{1}{\cal A} 
\left[ 
 \ln \left( {\cal A} \, x_{min} + {\cal B} \right) - \ln \left( {\cal C} \right)
\right] 
\label{interm2}
\end{eqnarray}
such that the residue of the fake pole $1/{\cal A}$ vanishes in each combination
$[\ln ({\cal A} + {\cal B} ) -  \ln ({\cal C} )] / {\cal A}$ and 
$\ln ({\cal A} \, x_{min} + {\cal B}) - \ln ({\cal C})] / {\cal A}$ separately.
The two terms in the r.h.s. of eq. (\ref{interm2}) thus lead to integrals
over $y^{\prime}$ which are individually well defined and may be safely handled 
on their own. A similar treatment may be done for $I_{3}^{6}$ adding and 
subtracting a term ${\cal C} \ln ({\cal C})$. We note that
\[
\left. {\cal B} \right|_{{\cal A} = 0} = - \,\frac{1}{2B} \, - \, i \lambda 
\] 
with 
\begin{eqnarray}
B  
& \equiv & 
\frac{\detg}{\dets}
\label{eeememeg}
\end{eqnarray}
thus we choose
\begin{eqnarray}
{\cal C} 
& = & 
- \,\frac{1}{2B} \, - \, i \lambda 
\label{eeememem}
\end{eqnarray}
In this way the integration over $x$ yields four terms. By means of an
appropriate change of variable, two of them may be further recombined so that 
each of the integrals $I^4_3$ and $I^6_3$ can be written as the sum of three 
terms. We call these terms ``sector integrals" labelled ${\cal I}_{(j)}$, 
$j=1,2,3$, they may be put in the following form. For $I^4_3$ we get:
\begin{eqnarray}
I^4_3 & = & \sum_{j=1}^{3} {\cal I}^{4}_{3 \, (j)}
\label{eqI37*}
\end{eqnarray}
with the sector integrals ${\cal I}^{4}_{3 \, (j)}$of the form
\begin{eqnarray}
{\cal I}^4_{3 \, (j)} 
& = &
- \, \int_{0}^{1} d z \, 
\frac{K_{(j)}(\alpha)}{D_{(j)} z + E_{(j)}}
\nonumber\\
&& \;\;\;\;\;\;\;\;
\left[ 
 \ln \left( F_{(j)} z^2 + G_{(j)} \, z + H_{(j)}- i \lambda \right)  
 \, - \, \ln \left( - \,\frac{1}{2B}  - i \lambda \right)
\right]
\label{eqI37**}
\end{eqnarray}
The coefficients $D_{(j)}, \cdots, K_{(j)}(\alpha)$ being provided by the 
following table; the dependence of the $K_{(j)}(\alpha)$ on $\alpha$ is made 
explicit for further convenience.
\begin{equation}\label{tablesectors}
\begin{array}{lll}
\mbox{sector $(1)$} & \mbox{sector $(2)$} & \mbox{sector $(3)$} \\
D_{(1)} = (2 b \, \alpha + c)  & D_{(2)} = (2 b \, \alpha + c)(-\alpha) & D_{(3)} =(2 b \, \alpha + c)(1-\alpha) \\
E_{(1)} = (d+ e \, \alpha)+(2a + c \, \alpha) & E_{(2)} = (d+ e \, \alpha) & E_{(3)} = (d+ e \, \alpha) \\
F_{(1)} =  b        & F_{(2)} = a & F_{(3)} =  (a+b+c)  \\
G_{(1)} =  (c+e)    & G_{(2)} = d & G_{(3)} =  (d+e)    \\
H_{(1)} = f+d+e  & H_{(2)} =  f & H_{(3)} =  f\\
K_{(1)}(\alpha) =  1 & K_{(2)}(\alpha) = - \, \alpha  & K_{(3)}(\alpha) = - \, (1-\alpha) 
\end{array}
\end{equation}
where $a,b, \cdots, f$ have been listed above in eq. (\ref{list}). 

\vspace{0.3cm}

\noindent
Similarly, for ${\cal I}^6_{3 \, (j)}$ we have: 
\begin{eqnarray}
I^6_3 & = & - \, \frac{1}{2} + \sum_{j=1}^{3} {\cal I}^6_{3 \, (j)}
\label{eqI37*}
\end{eqnarray}
with ${\cal I}^6_{3 \, (j)}$ of the form
\begin{eqnarray}
{\cal I}^{6}_{3 \, (j)}
& = &
\int_{0}^{1} d z \, 
\frac{K_{(j)}(\alpha)}{D_{(j)} z + E_{(j)}}
\nonumber\\
&& 
\mbox{} \times \Biggl[ 
 \left( F_{(j)} z^2 + G_{(j)} \, z + H_{(j)} \right) \, 
 \ln \left( F_{(j)} z^2 + G_{(j)} \, z + H_{(j)} - i \, \lambda \right) 
\nonumber\\
&& 
 \;\;\;\;\;\;\;\;\;\;\;\;\;\;\;\;\;\;\;\;\;\;\;\;\;\;\;\;\;\;\;\;\;\;\;
 \, + \,\frac{1}{2B}  \, \ln \left( - \,\frac{1}{2B}  - i \lambda \right)
\Biggr]
\label{eqI367**}
\end{eqnarray}
with $D_{(j)}, \cdots, K_{(j)}(\alpha)$ given in table (\ref{tablesectors}) above.

\noindent
The values of the integrals $I_{3}^{4}$ and $I_{3}^{6}$ do not depend on the 
particular root $\alpha = \alpha_{\pm}$ of eq. (\ref{eqTRINOME-ALPHA}) chosen 
to perform the first integration leading to eqs. (\ref{eqI37**}), 
(\ref{eqI367**}). As in ref. \cite{hv}, either of the two 
$\alpha$ roots, say $\alpha_{+}$, may be used to further compute the remaining 
single integrals in closed form in terms of logarithms and dilogarithms. 
A symmetrization over $\alpha_{\pm}$ would generate an unnecessary doubling of
dilogarithms in the closed form that would be prejudicial regarding CPU time 
in practice. However the discussion of the behaviours of these integrals
when $\detg \rightarrow 0$ is made somewhat obscure once one particular 
choice is made, and for this purpose it is on the contrary more enlightening to 
symmetrize expressions (\ref{eqI37**}) and (\ref{eqI367**}) over $\alpha_{\pm}$, 
especially in the perspective of providing one dimensional integral 
representations free of $\detg$ instabilities. The $\alpha$ dependence
comes only from the factors $K_{(j)}(\alpha)/(D_{(j)} z + E_{(j)})$, not from 
the arguments of the logarithms in numerators. Each of the sector integrals in 
the decomposition of $I_3^4$, respectively $I_3^6$, has an explicit $\alpha$ 
dependence of the type:
\[
{\cal I} = \int_0^1 dy \, \frac{K(\alpha)}{\alpha \, A + C} \, L
\]
where $L$ stands for the $\alpha$-independent numerators in the integrands of
the sector integrals ${\cal I}^{4,6}_{3 \, (j)}$, and we omit the superscript
$(j)$ labelling the sector for simplicity. Symmetrizing over $\alpha_{\pm}$ we 
get:
\begin{eqnarray}
{\cal I} 
& = & 
\frac{1}{2} \, \int_0^1 dy \, 
\left[ 
 \frac{(K(\alpha_{+}) \, \alpha_{-} + K(\alpha_{-}) \, \alpha_{+}) \, A + 
       (K(\alpha_{+}) + K(\alpha_{-}) ) \, C}{\alpha_{+} \, \alpha_{-} \, A^2 + 
        A \, C \, (\alpha_{-} + \alpha_{+}) + C^2}  
\right] \, L \nonumber
\label{eqALPHADEP}
\end{eqnarray}
Let us introduce the following quantities:
\begin{eqnarray*}
Q 
& = & 
\alpha_{+} \, \alpha_{-} \, A^2 + A \, C \, 
(\alpha_{-} + \alpha_{+}) + C^2 \\
& = & 
\frac{1}{b} \, ( a \, A^2 - c \, A \, C + b \, C^2) \\
N 
& = & 
(K(\alpha_{+}) \, \alpha_{-} + K(\alpha_{-}) \, 
\alpha_{+}) \, A + (K(\alpha_{+}) + K(\alpha_{-}) ) \, C
\end{eqnarray*}
Here are the explicit forms corresponding to the different sector integrals.\\
For sector $(1)$, $K(\alpha) = 1$, $A = 2 \, b \, z + e + c$, 
$C = c \, z + d + 2 \, a$, and we get :
\begin{eqnarray}
Q 
& = & 
\frac{1}{b} \, 
\left[ 
 - \deltal \, b \, z^2  - \deltal \, ( c+e) \, z 
 + a \, e^2 - c \, e \, d + b \, d^2 -  \deltal \, (d + a) 
\right] 
\nonumber \\
& = & 
\frac{1}{b} \, 
\left[
 \detg \, g_{(1)}(z)  + \frac{1}{2}\dets 
\right] 
\label{eqQ1} \\
N & = & 
\frac{1}{b} \, \left[ 2 \, b \, d - c \, e - \deltal \right] 
\nonumber \\
& = & \frac{1}{b} \, b_1 \, \dets
\label{eqN1}
\end{eqnarray}
For sector $(2)$, similarly,  
$K(\alpha) = - \, \alpha$, $A = c \, z + e$ and 
$C = 2 \, a \, z + d$, so that :
\begin{eqnarray}
Q 
& = & 
\frac{1}{b} \, 
\left[ 
 - \deltal \, a \, z^2 - \deltal \, d \, z 
 + a \, e^2 - c \, e \, d + b \, d^2 
\right] 
\nonumber \\
& = & 
\frac{1}{b} \, 
\left[ \detg \, g_{(3)}(z) + \frac{1}{2} \dets
\right] 
\label{eqQ3} \\
N & = & - \, \frac{1}{b} \, \left[ 2 \, a \, e - c \, d \right] \nonumber \\
& = & \frac{1}{b} \, b_2 \, \dets
\label{eqN3}
\end{eqnarray}
For sector $(3)$, 
$K(\alpha) = - \, (1-\alpha)$, $A = (2 \, b + c) \, z + e$ and 
$C = (c + 2 \, a) \, z + d$, so that:
\begin{eqnarray}
Q 
& = & 
\frac{1}{b} \, 
\left[ 
 - \deltal \, (a+b+c) \, z^2 
 - \deltal \, ( e+d) \, z + a \, e^2 - c \, e \, d + b \, d^2 
\right] 
\nonumber \\
& = & 
\frac{1}{b} \, 
\left[ 
 \detg \, g_{(2)}(z) + \frac{1}{2} \, \dets 
\right] 
\label{eqQ2} \\
N 
& = & 
- \, \frac{1}{b} \, 
\left[ 
 2 \, b \, d  + c \, d - 2 \, a \, e - c \, e 
\right] 
\nonumber \\
& = & 
\frac{1}{b} \, b_3 \, \dets
\label{eqN2}
\end{eqnarray}
where the coefficients $b_{j}$ are defined by
\begin{equation}\label{defbj}
b_j = \sum_{k=1}^{3} \, \cals^{-1}_{j \, k}
\end{equation}
They are such that
\begin{equation}\label{propbj}
\sum_{j=1}^{3} \, b_j  = B = \frac{\detg}{\dets}
\end{equation}
They were introduced in the {\tt GOLEM} reduction algorithm \cite{golem},
and the second degree polynomials $g_{(j)}(z)$ are given by
\begin{equation}\label{gjzdef}
\begin{array}{ccl}
g_{(1)}(z) & = & b \, z^2 + (c+e) \, z + (a + d +f) \\
g_{(2)}(z) & = & a \, z^2 + d \, z + f \\
g_{(3)}(z) & = & (a+b+c) \, z^2 + (d+e) \, z +f 
\end{array}
\end{equation}
The polynomials $g_{(j)}(z)$ are namely those appearing in the integral 
representations of the two-point functions corresponding to the three possible 
pinchings of one propagator in the triangle diagram of Fig. 1. 
In what follows we parametrize the $g_{(j)}(z)$ generically as
\begin{equation}\label{gj-param}
g_{(j)}(z) = 
\gamma_{(j)}^{\prime \prime} \, z^2 + \gamma_{(j)}^{\prime} \, z + \gamma_{(j)} 
\end{equation}
in order to formally handle them all at once when concerned with the zeroes 
of $g_{(j)}(z) + 1/(2B)$ further below. Let us note that the discriminant 
$\Delta_{j}$ of the second degree polynomial $g_{(j)}(z)$, defined by 
\begin{eqnarray}\label{deltaj}
\Delta_{j} 
& \equiv & 
\gamma_{(j)}^{\prime \, 2} - 4 \, \gamma_{(j)}^{\prime \prime} \, \gamma_{(j)}
\end{eqnarray}
turns out to be equal to minus the determinant of the reduced kinematic matrix 
$\cals^{\{j\}}$. This reduced kinematic matrix corresponds to the 
pinching of the propagator $j$ in the triangle of Fig. \ref{fig1}, and is obtained from
the matrix $\cals$ by suppressing line and column $j$. Correlatively 
$\gamma_{(j)}^{\prime \prime}$ can be seen as half the
reduced Gram determinant associated with the reduced kinematic matrix 
$\cals^{\{j\}}$.
\vspace{0.3cm}

\noindent
Equation (\ref{eqI37**}) can thus be written:
\begin{eqnarray}
I^4_3 
& = &
- \, \sum_{j=1}^{3}
b_{j} 
\int_{0}^{1} d z \, 
\frac{\ln \left( g_{(j)}(z) \right) - \ln \left( -1/(2 \, B) \right)}
{2 \, B \, g_{(j)}(z)+1} 
\label{eqI38}
\end{eqnarray}
Likewise for eq. (\ref{eqI367**}):
\begin{eqnarray}
I^6_3 
& = & 
- \, \frac{1}{2} + 
\sum_{j=1}^{3} 
b_{j} \, 
\int_{0}^{1} d z \, 
\frac{g_{(j)}(z) \, \ln \left( g_{(j)}(z) \right) + 
1/(2 \, B) \, \ln \left( -1/(2 \, B) \right)}
{2 \, B \, g_{(j)}(z)+1}   
\label{eqI368}
\end{eqnarray}
In eqs. (\ref{eqI38}), (\ref{eqI368}), the contour prescription inherited from 
$(- \, z^{T} \cals z - i \lambda)$ in eqs. (\ref{eqSTARTINGPOINT}),
(\ref{eqSTARTINGPOINT6}) is implicit: the logarithmic terms 
$\ln \left( g_{(j)}(z) \right)$ in the numerators stand for 
$\ln \left( g_{(j)}(z) - i \lambda \right)$. Let us remind that the terms 
$\ln (- 1/(2B))$ in the numerators have been introduced in
order that the zeroes $z_{(j)}^{\pm}$ of the denominators 
$(2B \, g_{(j)}(z) +1)$
be  fictitious poles in each of the sector integrals  in any case i.e. the
residues vanish: hence $\ln (- 1/(2B))$ stands for $\ln (- 1/(2B)- i \lambda)$
as well, furthermore no contour prescription around the $z_{(j)}^{\pm}$ is
needed. 

\vspace{0.3cm}

\noindent
Equations (\ref{eqI38}) and (\ref{eqI368}) are appealing candidates for the 
integral representations which we seek. 
Let us examine them more closely when $\detg \to 0$. 
We shall distinguish two cases: the generic case when $\detg \to 0$ whereas 
$\dets$ remains non vanishing, and the specific case $\detg \to 0$ and 
$\dets \to 0$ simultaneously which deserves a dedicated treatment. 
Let us subsequently examine these two cases.

\mysection{$\detg \to 0$ whereas $\dets$ non vanishing}\label{sect3}

Let us first consider the polynomials $g_{(j)}(z)+1/(2B)$ appearing in the 
denominators of the integrals ${\cal I}^{4,6}_{3 \, (j)}$ in eqs. 
(\ref{eqI38}), (\ref{eqI368}). Let us first consider
$\gamma_{(j)}^{\prime \prime} \neq 0$, so that $g_{(j)}(z) +1/(2 B)$ is 
of degree two. Using the identity
\begin{equation}
  \bbar_j^2 = 2 \, \gamma_{(j)}^{\prime \prime} \, \dets - \detg \, \Delta_j
  \label{eqI41*}
\end{equation}
where $\Delta_j$ has been defined in eq. (\ref{deltaj}), and the rescaled 
coefficients
\begin{equation}\label{rescaledbj}
\bbar_j \equiv b_{j} \dets, \;\;\;\; j=1,2,3
\end{equation}
it is insightful to write the corresponding 
discriminant of $g_{(j)}(z)+1/(2B)$ as 
\begin{eqnarray}
\widetilde{\Delta}_{j} 
& = & 
-  \, \frac{\bbar_j^2}{\detg}
\label{eqI40*}
\end{eqnarray}
Identity (\ref{eqI41*}) is derived in Appendix \ref{algebraic-identities}. 
It is an example of the so-called Jacobi identities for determinant ratios,
relating the determinant of a matrix and related cofactors i.e. determinants of 
reduced matrices \cite{encyclopedymath,fbl}. Similar identities 
may be met in the treatment of the four-point function. 
The zeroes $z_{(j)}^{\pm}$ of $g_{(j)}(z)+1/(2B)$ are given by 
\begin{eqnarray}
z_{(j)}^{\pm} 
& = & 
 - \frac{\gamma_{(j)}^{\prime}}{2 \, \gamma_{(j)}^{\prime \prime}} 
\mp \frac{\bbar_j}{2 \, \gamma_{(j)}^{\prime \prime}\, \sqrt{-\detg}}
\label{zjpm}
\end{eqnarray}
(as commented earlier, $\detg \leq 0$).
When $\detg \to 0$, both zeroes $z_{j}^{\pm}$ of $2B g_{(j)}(z)+1$
are dragged away from $[0,1]$ towards $+ \, \infty$ and $- \, \infty$
respectively. If $\gamma_{(j)}^{\prime \prime} = 0$, $g_{(j)}(z) +1/(2 B)$ is 
only of degree one, and its unique root $z_{(j)}^{0}$ given by
\begin{eqnarray}
z_{(j)}^{0} 
& = & 
 - \, \frac{1}{\gamma_{(j)}^{\prime}} 
\left( 
 \gamma_{(j)} + \frac{1}{2} \, \frac{\dets}{\detg} 
\right)
\label{zj0}
\end{eqnarray}
is again dragged away from $[0,1]$ towards $\infty$ when $\detg \to 0$. 
In either case, as soon as $\detg$ becomes small enough each of the 
integrals
\[
{\cal J}_{j} = \int_{0}^{1} \frac{dz}{2B g_{(j)}(z)+1} 
\]
is analytically well defined and numerically safe, and furthermore the following
identity holds:
\begin{equation}\label{fictituouscancel}
\sum_{j=1}^{3} b_{j} {\cal J}_{j} = 0
\end{equation} 
so that the contributions $\propto \ln( - 1/(2B) - i\lambda)$ sum up to zero 
in $I_3^4$ as well as in $I_3^6$. In this respect, let us stress that the 
contributions $\propto \ln( - 1/(2B) - i\lambda)$ are fictitious from the 
start. 
They were introduced through eq. (\ref{interm2}) with the custodial concern of 
separately handling integrals - the sector integrals - with integrands free of 
poles within the integration domain namely when either of $z_{j}^{\pm}$ is 
inside $[0,1]$. When $z_{j}^{\pm}$ are both outside $[0,1]$ the introduction of 
the $\ln( - 1/(2B) - i\lambda)$ terms is irrelevant and indeed identity 
(\ref{fictituouscancel}) allows to drop them explicitly from eqs. 
(\ref{eqI38}),(\ref{eqI368}). The following integrals
\begin{eqnarray}
I^4_3 
& = &
- \, \sum_{j=1}^{3}
b_{j} 
\int_{0}^{1} d z \, 
\frac{\ln \left( g_{(j)}(z) - i \lambda\right)}{2 \, B \, g_{(j)}(z)+1} 
\label{eqI38s}
\\
I^6_3 
& = & 
- \, \frac{1}{2} + 
\sum_{j=1}^{3} 
b_{j} \, 
\int_{0}^{1} d z \, 
\frac{g_{(j)}(z) \, \ln \left( g_{(j)}(z) - i \, \lambda \right)}
{2 \, B \, g_{(j)}(z)+1}   
\label{eqI368s}
\end{eqnarray}
thus provide suitable integral representations in the case at hand. 
From a numerical point of view the explicit suppression of the
$\ln( - 1/(2B) - i\lambda)$ terms from integrals (\ref{eqI38s}), 
(\ref{eqI368s}) is preferable since $\ln( - 1/(2B) - i\lambda)  \to \infty$ 
when $\detg \to 0$ thus implementing a numerical cancellation of the sum 
$\sum_{j=1}^{3} b_{j} {\cal J}_{j} \ln( - 1/(2B) - i\lambda)$ after each term 
would have been separately calculated would be submitted to numerical 
instabilities. Besides, if case some $g_{(j)}(z)$ vanishes at some 
$\hat{z}_{(j)}$ inside $[0,1]$, a possible numerical improvement of the
integral representation  consists in deforming the integration contour in the
complex $z$ plane, to skirt the vicinity of the integrable singularity at
$\hat{z}_{(j)}$, so as to prevent the integrand from becoming large and
avoid cancellation of large contributions, according to a one-dimensional
version\footnote{In broad outline, the contour deformation is contained inside
the band $0 \leq {\cal R}e(z) \leq 1$. It departs from the real axis at 0 with
an acute angle and likewise ends at 1 in such a way that  ${\cal I}m(g_{(j)}(z))$ 
is kept negative along
the deformed contour so that the latter does not cross any cut of $\ln
(g_{(j)}(z) - i \, \lambda)$. In the case at hand this type of contour never
embraces any of $z_{j}^{\pm}$ as soon as the latter are outside $[0,1]$, thus no
subtraction of illegitimate pole residue contribution at $z_{j}^{\pm}$ has to be
cared about.}  of the multidimensional deformation described in section 7 of
ref. \cite{golem}.

\mysection{$\detg \to 0$ and $\dets \to 0$ simultaneously}\label{sect4}

This case is more tricky and deserves further discussion. Indeed, when 
$\dets =0$ and $\detg = 0$, eq. (\ref{defbj}) defining the parameters 
$b_{j}$ as $\sum_{k=1}^{3} \cals^{-1}_{jk}$ is no longer valid as 
$\cals^{-1}$ is not defined, and the parameter $B = \dets/\detg$ is an 
indeterminate quantity of the type 0/0, likewise the $z_{(j)}^{\pm}$ are 
indeterminate quantities not manifestly driven away from the interval $[0,1]$. 

\vspace{0.3cm}

\noindent
In this subsection we will first characterize the specific kinematics which 
leads to such a case. Then we will consider kinematic configurations close to 
the so-called specific ones above, such that $\detg$ and $\dets$ are 
simultaneously small but non vanishing and we will study how
$I_{3}^{4}$ and $I_{3}^{6}$ behave when $\dets$ and  $\detg$ both go to
zero. Anticipating on the result, we rewrite both for $I_{3}^{4}$ and 
$I_{3}^{6}$ the corresponding sums
\begin{equation}\label{recombI3}
\sum_{j=1}^{3} b_{j} \, {\cal I}_{3 \, (j)} 
= b_{3} \, {\cal I}_{3 \, (3)} + 
\frac{1}{2} 
(b_{1} + b_{2}) \left( {\cal I}_{3 \, (1)}+{\cal I}_{3 \, (2)}\right) +
\frac{1}{2} 
(b_{1} - b_{2}) \left( {\cal I}_{3 \, (1)}-{\cal I}_{3 \, (2)}\right)
\end{equation}
One of the coefficients $b_{j}$, say $b_{3}$ will be shown to have a finite 
limit whereas $b_{1}$ and $b_{2}$ diverge towards infinity in a concomitant way 
such that their sum $b_{1}+b_{2}$ has a finite limit. 
Furthermore, the difference ${\cal I}_{3 \, (1)}-{\cal I}_{3 \, (2)}$ will be 
shown to tend to zero so that the product 
$(b_{1}-b_{2})({\cal I}_{3 \, (1)}-{\cal I}_{3 \, (2)})$ has a finite limit.  
A well-defined expression is thus achieved in the double limit 
$\dets \to 0$, $\detg \to 0$ although some of the ingredients are
separately ill-defined in the limit considered. We will conclude this subsection
with a comment in relation with the behaviour of the {\tt GOLEM}
reduction formalism in this case.

\subsection{Characterization of the specific kinematics 
$\detg = 0$, $\dets = 0$}\label{sss221}
  
The quantities $\dets$ and $\detg$ are polynomials in the kinematical 
invariants. Hereafter we propose a presentation which partly linearizes the 
resolution of the non linear system $\detg = 0$, $\dets = 0$. This
approach, applied here to $N=3$, extends to other $N$, e.g. $N=4$. 
The determinant $\dets$ can be written (see Appendix 
\ref{algebraic-identities}):
\begin{equation}\label{detsdetggtilde}
\dets =
{\cal S}_{aa} \, \detg + V^{(a) \, T} \cdot \widetilde{G}^{(a)} \cdot V^{(a)}
\end{equation}
where $\widetilde{G}^{(a)}$ is the matrix of cofactors\footnote{This matrix is 
sometimes also called `adjoint matrix' of $G^{(a)}$.} of $G^{(a)}$; the 
superscript ``$^{T}$" refers to matrix transposition. 
The system $\detg = 0$, $\dets = 0$ is thus equivalent to the system
$\detg = 0$, $V^{(a) \, T} \cdot \widetilde{G}^{(a)} \cdot V^{(a)} = 0$. Since
\begin{equation}\label{ggtilde}
G^{(a)} \, \tilde{G}^{(a)} = \tilde{G}^{(a)}  \, G^{(a)} 
= 
\detg  \, 1\! \mbox{I}_{N}
\end{equation}
the matrices $G^{(a)}$ and $\tilde{G}^{(a)}$ are simultaneously diagonalizable.
When $\detg = 0$ and $G^{(a)}$ has rank\footnote{See comment at the beginning of
appendix \ref{app-degen} regarding $(N-1) \times(N-1)$ Gram matrices of lower 
ranks. In the present $N=3$ case we discard the degenerate 
possibility that the $2 \times 2$  matrix $G^{(a)}$ has two vanishing eigenvalues 
which not only makes the cofactor matrix $\widetilde{G}^{(a)}$ vanish
identically but also $G^{(a)}$ itself. This would correspond to a very peculiar 
kinematics of three lightlike external four-momenta collinear to each other.} 
$(N-2)$ - namely 1 in the $N=3$ case at hand - the only eigendirection
$\hat{n}^{(a)}$ of $G^{(a)}$ associated to the eigenvalue zero is 
concomitantly the only eigendirection of $\tilde{G}^{(a)}$ 
associated to the only non vanishing eigenvalue $\tilde{g}$ of 
$\tilde{G}^{(a)}$. For $\hat{n}^{(a)}$ properly normalized,
$\tilde{G}^{(a)} = \tilde{g} \; \hat{n}^{(a)} \otimes \hat{n}^{(a) \, T}$. The 
condition 
$V^{(a) \, T} \cdot \tilde{G}^{(a)} \cdot V^{(a)} = 0$  quadratic in $V^{(a)}$
is thus equivalent to the following linear one:
\begin{equation}\label{linearized}
( \hat{n}^{(a) \, T} \cdot V^{(a)}) = 0
\end{equation}
Let us now consider the condition $\detg = 0$. A detailed discussion is 
provided in appendix \ref{detg-zero-kinematics}, we
only summarize it here for the $N=3$ case at hand. A vanishing $\detg$
happens (i) either when the external momenta $p_{1,2,3}$ are proportional to 
each other (ii) or when there exists a non vanishing linear combination of the 
external momenta which is lightlike and orthogonal to all of them \cite{kinem}.
Possibility (i) corresponds to degenerate kinematic configurations irrelevant 
for next-to-leading order (NLO) calculations of collider processes. Let us 
focus on possibility (ii) further assuming any subset of two of the three 
external momenta to be linearly independent. To fix the ideas, let us 
consider\footnote{This particular choice corresponds to singling out and 
erasing line and column 3 in the matrix ${\cal S}$ and considering the 
Gram matrix $G^{(3)}$.} $p_{1}$ and $p_{3}$. If one of them say 
$p_{1}$ is lightlike it is namely (proportional to) the lightlike combination 
sought, whereas $p_{3}$ shall be spacelike, $p_{2}= -p_{1}-p_{3}$ is 
spacelike as well and $s_{2} = s_{3}$. If neither $p_{1}$ nor $p_{3}$ are 
lightlike, both shall be spacelike with $s_{1} = s_{3}$, and $p_{2}$ is 
(proportional to) the lightlike combination of $p_{1}$ and $p_{3}$.
Actually, configurations of type (ii) with $p_{3}$, $p_{1}$ and $p_{2}$ 
linearly independent and all spacelike can also lead to a vanishing $\detg$, 
yet such configurations are not  relevant for collider processes at
NLO\footnote{Indeed, at NLO, each of the external legs of the one loop three
point function considered has to be connected to separate {\em tree}, and all
the external legs of at least one of these three trees have to be final state
legs. Therefore the external momentum flowing through the corresponding leg of
the one loop three point function cannot be spacelike. 
Such configurations with three
spacelike legs could appear only in higher loop diagrams, of which the one loop
three point function would be seen as a subdiagram.}, we thus discard them. 

\vspace{0.3cm}

\noindent
Let us assume $p_{2}$ lightlike and orthogonal to $p_{3}$ and $p_{1}$ both 
spacelike: $s_{1} = s_{3} \equiv s_{+} < 0$, $s_{2} = 
(p_{2} \cdot p_{3,1}) = 0$, so that $(p_{1} \cdot p_{3}) = - \, s_{+}$. 
We single out line and column 3 of ${\cal S}$ whose corresponding 
$G^{(3)}$ reads:
\begin{equation}\label{G3}
G^{(3)} 
= 
2 \, s_{+}
\left[
 \begin{array}{cc}
 1 & 1 \\
 1 & 1
 \end{array}
\right]
\end{equation}
The normalized eigenvector $\hat{n}^{(3)}$ associated with the eigenvalue zero 
is (up to a sign):
\begin{equation}\label{u}
\hat{n}^{(3)} 
= 
\frac{1}{\sqrt{2}} \,
\left[
 \begin{array}{r}
 1  \\
 - \, 1
 \end{array}
\right]
\end{equation}
With eqs. (\ref{eqDEFV3}) and (\ref{u}), condition (\ref{linearized})
imposes the following restriction on the internal masses:
\begin{equation}\label{cond}
m_{1}^{2} = m_{2}^{2} \equiv m^{2}
\end{equation}

\subsection{Behaviour of $I_{3}^{4}$ and $I_{3}^{6}$ when $\detg \to 0$, 
$\dets \to 0$}\label{ss222}

Let us assume condition (\ref{cond}) and parametrize the 
departure from the `critical kinematics' using $s_{2}$,
$s_{-} \equiv (s_{1} - s_{3})/2$ and $s_{+} \equiv  (s_{1} + s_{3})/2$. 
The determinants read:
\begin{eqnarray}
\detg 
& = & 
4 \, 
\left( 
 s_{+} \, s_{2} \, - \, s_{-}^{2} - \frac{1}{4} s_{2}^{2}
\right)
\label{approxdetg}\\
\dets 
& = &
2 \, \left( 
 \widetilde{\lambda} \, s_{2} + 4 \, m^{2} \, s_{-}^{2} \, + \, 
 m_{3}^{2} \, s_{2}^{2} \, - \, s_{2} \, s_{-}^{2}
\right)
\label{approxdetS}
\end{eqnarray}
where $\widetilde{\lambda}$ is the K\"allen symmetric function of 
$s_{+},m^{2},m_{3}^{2}$ given by:  
\begin{eqnarray}
\widetilde{\lambda} 
& = &
s_{+}^{2} + (m^{2})^{2} + (m_{3}^{2})^{2} - 2 \, m^{2} \, s_{+}
- 2 \, m_{3}^{2} \, s_{+} - 2\, m^{2} \, m_{3}^{2} \; > \; 0
\label{lambda}
\end{eqnarray}
The region where $\detg$ and $\dets$ are concomitantly 
small corresponds to $|s_{2}|$, $|s_{-}|$ both small compared with 
the other kinematical invariants, with $|s_{2}/s_{+}|$ and 
$(s_{-}/s_{+})^{2}$ of the same order, so that $\detg$ and 
$\dets$ can be approximated by: 
\begin{eqnarray}
\detg 
& = & 
4 \, 
\left(  s_{+} \, s_{2} \, - \, s_{-}^{2} \right) 
+ \cdots
\label{approxdetg2}\\
\dets 
& = &
2 \, \left(  \widetilde{\lambda} \, s_{2}+ 4 \, m^{2} \, s_{-}^{2} \right)
+ \cdots
\label{approxdetS2}
\end{eqnarray} 
In order to understand in more detail the origin of the diverging contributions
to the coefficients $b_{j}$, the matrix ${\cal S}$ may be decomposed as follows:
\begin{equation}\label{diag-S}
{\cal S} = \sum_{j=1}^{3} \sigma_{j} \, v_{(j)} \otimes v_{(j)}^{T} 
\end{equation}
Let us address the real mass case first; we will briefly comment at the end of 
this subsubsection on how the study shall be - only slightly - modified 
in the complex mass case. Decomposition (\ref{diag-S}) 
corresponds to the usual diagonalization of ${\cal S}$: the $v_{(j)}$ and 
$\sigma_{j}$, $j =1,2,3$ are real eigenvectors orthonormalized in the 
euclidean sense $(v_{(j)}^{T} \cdot v_{(k)}) = \delta_{jk}$ and the
corresponding real eigenvalues respectively. The labelling of eigenvectors and
values is chosen such that $\sigma_{i=3}$ explicitly given by
\begin{equation}\label{sig3}
\sigma_{3} 
= 
- \, \left( s_{2} + \frac{4 \, m^{2}}{\widetilde{\lambda}} \, s_{-}^{2} \right)
+ \cdots 
\end{equation}
is the eigenvalue which vanishes when $s_{2}$ and $s_{-}$ both 
vanish, whereas the two others remain finite in this limit. Introducing
\begin{equation}\label{def-e}
e 
\equiv  
\left[
 \begin{array}{c}
 1  \\
 1  \\
 1
 \end{array}
\right]
\end{equation}
the column vector $b = {\cal S}^{-1} \cdot e$ and the quantity 
$B = \sum_{j=1}^{3} b_{j} = (e^{T}.b)$ take the form:
\begin{eqnarray}
b 
& = &
\sum_{i=1}^{3} \sigma_{i}^{-1} \, \left(v_{(i)}^{T} \cdot e \right) \, v_{(i)} 
\label{bj-1}\\
B 
& = &
\sum_{i=1}^{3} \sigma_{i}^{-1} \, \left(v_{(i)}^{T} \cdot e \right) ^{2} 
\label{B-1}
\end{eqnarray}
More explicit  algebraic expressions of the various
ingredients in the relevant regime are gathered in appendices
\ref{spectral-decomp-S} and \ref{calcbj-B} for convenience.
They show that $(v_{(i)}^{T} \cdot e )\sim {\cal O}(s_{-})$ 
so that the components of  
$b = {\cal S}^{-1} \cdot e$ are individually wild-behaved when 
$\dets \to 0$ due to the ${\cal O}(s_{-}^{2}/\sigma_{3})$ 
contribution along $v_{(3)}$ being  ${\cal O}(|\sigma_{3}|^{-1/2})$, 
although $B \sim {\cal O}(s_{-}^{2}/\sigma_{3})$ remains  
${\cal O}(1)$. A closer look reveals that both $b_{3} \sim
{\cal O}(s_{2}/\sigma_{3})$ and the combination 
$(b_{1}+b_{2}) \sim {\cal O}(s_{2}/\sigma_{3}, s_{-}^{2}/\sigma_{3})$ 
 separately remain  ${\cal O}(1)$ whereas 
$(b_{1}-b_{2}) \sim {\cal O}(s_{-}/\sigma_{3})$ is  
${\cal O}(|\sigma_{3}|^{-1/2})$. Concomitantly, since
\begin{eqnarray}
g_{(1)}(z) & = & g(z) + s_{-} \, z \, (1-z)
\label{limg1}\\
g_{(2)}(z) & = & g(1-z) - s_{-} \, z \, (1-z) 
\label{limg2}
\end{eqnarray}
where
\begin{eqnarray}
g(z) 
& = & - \, s_{+} \, z \, (1 - z) + m^{2} \, z + m_{3}^{2} \, (1-z)
\label{limg}
\end{eqnarray}
the quadratic forms $g_{(1)}(z)$ and $g_{(2)}(1-z)$ become both equal to 
$g(z)$ when $s_{-}= 0$. The difference of 
the two integrals ${\cal I}_{3 \, (1)}$ and ${\cal I}_{3 \, (2)}$ in factor of 
$b_{1}$ and $b_{2}$ respectively, in eqs. (\ref{eqI38}) for $I_{3}^{4}$ 
and likewise in (\ref{eqI368}) for $I_{3}^{6}$, is 
$({\cal I}_{3 \, (1)} - {\cal I}_{3 \, (2)}) \sim {\cal O}(s_{-})$. 
The combination 
$(b_{1}-b_{2}) \, ({\cal I}_{3 \, (1)} - {\cal I}_{3 \, (2)})$ is thus
$\sim ({\cal O}(s_{-}^{2}/\sigma_{3})$ i.e  ${\cal O}(1)$ as 
well. In the summary, rewriting $\sum_{j=1}^{3} b_{j} {\cal I}_{3 \, (j)}$
according to eq. (\ref{recombI3}), each of the three terms 
$b_{3} \, {\cal I}_{3 \, (j)}$, 
$(b_{1}+b_{2}) \, ({\cal I}_{3 \, (1)} + {\cal I}_{3 \, (2)})$ and
$(b_{1}-b_{2}) \, ({\cal I}_{3 \, (1)} - {\cal I}_{3 \, (2)})$
remains bounded and has a finite limit when $s_{-} \to 0$, 
$s_{2} \to 0$.

\vspace{0.3cm}

\noindent
Let us however notice that we are taking a {\em double limit}. Properly 
speaking, the limits of each of these three terms in eq. 
(\ref{recombI3}) which are separately well-defined are {\em directional limits} 
$s_{-} \to 0$, $s_{2} \to 0$ in the 
$\{ s_{-}^{2}, s_{2}\}$ plane keeping the ratio 
$t = s_{2}/s_{-}^{2}$ fixed, i.e. these directional limits are
functions of $t$. However, the limit of the {\em sum} of 
these three terms in eq. (\ref{recombI3}) is indeed {\em independent} of $t$. 
This can be easily checked numerically, this can also be proven analytically
although this is somewhat cumbersome; a proof is presented in Appendix
\ref{appendix_b}. 
The ground reason why this property holds is further understood as follows: 
were the limit of the sum a directional one, it would imply that the 
three-point 
function would be a singular i.e. non analytical function of the 
kinematical invariants at such configurations. However the kinematic 
singularities are characterized by the so-called Landau conditions\footnote{For 
general parametric integrals the Landau  conditions provide only necessary
conditions to face singularities, either of  pinched or end-point type. However
for Feynman integrals, Coleman and Norton  \cite{coleman-norton} proved these
conditions to be sufficient.} \cite{kinem,landaucond} (see also \cite{fbl}). For one
loop diagrams, these conditions require not only that $\dets =0$, but also 
that the eigenvectors associated with the vanishing eigenvalue of ${\cal S}$ 
shall have only non negative components and that their sum be strictly positive. 
By contrast, in the case at hand, the eigenvector $v_{(3)}$ in the limit where
$\sigma_{3} = 0$  is, cf. Appendix \ref{spectral-decomp-S}:
\begin{equation}\label{v3lim}
v_{(3)} |_{\sigma_{3} = 0} 
\propto
\left[
 \begin{array}{r}
 1  \\
 - \, 1  \\
 0
 \end{array}
\right]
\end{equation}
such that $(e^{T} \cdot  v_{(3)}|_{\sigma_{3} = 0}) = 0$. The vanishing
$\dets$ in the present case is therefore {\em not} related to a 
kinematic singularity: the three-point function is {\em regular} in the limit
considered, in particular this limit shall be uniform i.e. not directional.

\subsection{Extension to the complex mass case}\label{ss223}

The above study was stressed to hold, strictly speaking, for real masses.
Actually, it can be extended to the complex mass case with only slight
modifications. Indeed in the complex mass case, the symmetric matrix ${\cal S}$
albeit complex admits a decomposition formally identical eq. (\ref{diag-S}):
\begin{equation}\label{diag-Scx}
{\cal S} = \sum_{j=1}^{3} \sigma^{\prime}_{i} \, u_{(j)} \otimes u_{(j)}^{T} 
\end{equation}
which now reflects the so-called Takagi factorization 
${\cal S} = U \cdot \Sigma \cdot U^{T}$ in terms of a real non negative 
diagonal matrix $\Sigma$ and a unitary matrix $U$, instead of a 
standard diagonalization. The diagonal elements $\sigma^{\prime}_{j}$ of
$\Sigma$ are the square roots of the eigenvalues of the hermician matrix 
${\cal S}\, {\cal S}^{\dagger}$, whereas the  columns $u_{(i)}$ of $U$ are
corresponding eigenvectors\footnote{Let us note  by passing that, unlike with
standard diagonalization, the phases of the vectors $u_{(i)}$ involved in the 
Takagi factorization shall be adjusted modulo $\pi$ in order to fulfill the 
condition $(u_{(j)}^{T} \cdot {\cal S} \cdot u_{(k)}^{*}) = 
\sigma_{j}^{\prime} \, (u_{(j)}^{T} \cdot u_{(k)})$, 
because the decomposition involves the transpose of $U$ not its hermician 
conjugate.} of ${\cal S} \, {\cal S}^{\dagger}$. The corresponding 
Takagi factorization of ${\cal S}^{-1}$ for ${\cal S}$ invertible reads 
${\cal S}^{-1} = U^{*} \cdot \Sigma^{-1} \cdot U^{\dagger}$ i.e. in the tensor 
product notation: 
\begin{equation}\label{diag-Sm1cx}
{\cal S}^{-1} = \sum_{i=1}^{3} 
\sigma^{\prime \, -1}_{j} \, u_{(j)}^{*} \otimes u_{(j)}^{\dagger} 
\end{equation}
Identity (\ref{diag-Sm1cx}) provides the equations for $b$ and $B$ which modify 
eqs. (\ref{bj-1}), (\ref{B-1}) in the complex mass case. A study  quite 
similar to the real mass case then follows\footnote{Technically speaking, the 
determination of the singular values $\sigma_{j}^{\prime}$ and corresponding
vectors $u_{(j)}$ may seem somewhat awkward given the algebraically more
complicated form of the matrix elements of ${\cal S} \, {\cal S}^{\dagger}$.
Actually we are interested in a practical case where the imaginary parts of the
masses - i.e widths of unstable particles in internal lines - are
much smaller than the real parts. Therefore, splitting  ${\cal S}$ in real and
imaginary parts ${\cal S} ={\cal S}_{R} - i \, {\cal S}_{I}$, and writing 
${\cal S} \, {\cal S}^{\dagger} ={\cal S}_{R}^{2} + \Delta$, with 
$\Delta = i[{\cal S}_{R},{\cal S}_{I}] + \, {\cal S}_{I}^{2}$, the square roots
$\sigma_{j}^{\prime}$ of eigenvalues of ${\cal S} \, {\cal S}^{\dagger}$ 
and the corresponding eigenvectors $u_{(j)}$ can be expanded in integer powers 
of matrix 
elements of ${\cal S}_{I}$, as perturbative deformations of the eigenvalues 
$\sigma_{k}$ and eigenvectors $v_{(k)}$ of ${\cal S}_{R}$ i.e. the spectral 
features of the real mass case, by a straightforward application of the formalism 
of time-independent perturbation theory in Quantum Mechanics.} and 
the same conclusions hold. 
 
\subsection{A comment on the {\tt GOLEM} reduction 
formalism \\when $\dets = 0$}\label{ss224}

Let us end this subsection with a comment on the applicability of the 
{\tt GOLEM} reduction formalism \cite{golem} to configurations such that 
$\dets =0$. The equation ${\cal S} \cdot b = e$ with ${\cal S}$ not 
invertible can still be solved e.g. introducing the so-called Moore-Penrose 
Pseudo-Inverse \cite{barnett} ${\cal T}_{0}$ of ${\cal S}(\sigma_{3} = 0)$ 
given
in the real mass case\footnote{A similar discussion holds in the complex mass 
case as well with similar expressions cf. the previous paragraph.} by: 
\begin{eqnarray}
{\cal S}_{\sigma_{3} = 0} 
& = &
\sum_{i=1,2} \sigma_{i} \, v_{(i)} \otimes v_{(i)}^{T}
\label{Ssing}\\
{\cal T}_{0} 
& = &
\sum_{i=1,2} \sigma_{i}^{-1} \, v_{(i)} \otimes v_{(i)}^{T}
\label{T}
\end{eqnarray}
{\em provided} the following compatibility condition to be satisfied:
\begin{equation}\label{compcond}
\left[ 1\! \mbox{I} -  \cals|_{\sigma_{3} = 0} \cdot {\cal T}_{0} \right] \cdot e
= 0
\end{equation}
Still noting $v_{(3)}$ the eigenvector of ${\cal S}$ with vanishing 
eigenvalue, the compatibility condition (\ref{compcond}) reads 
\begin{equation}\label{compcond2}
\left( e^{T} \cdot v_{(3)} \right) = 0
\end{equation}
Condition (\ref{compcond2}) is incompatible with the Landau conditions mentioned 
earlier which characterizes a kinematic singularity, namely the non negativity 
of all the components of $v_{(3)}$: thus the formalism breaks down for singular 
kinematics. 

\vspace{0.3cm}

\noindent
On the other hand, the ``peculiar" configurations such that 
$\detg = 0$, $\dets = 0$ 
examined in the present subsection are non singular and do fulfill condition 
(\ref{compcond2}), and the (non unique)\footnote{The arbitrary component of $b$ 
along $v_{(3)}|_{\sigma_{3} = 0}$ is irrelevant for any practical 
purpose. Indeed, the condition (\ref{compcond2}) makes the contribution to
$b_{3}$, hence to $B$, coming from this component vanish, whereas the finite
contribution to $b_{1}-b_{2}$ from this component is weighted by the vanishing 
${\cal I}_{3 \, (1)} - {\cal I}_{3 \, (2)}$ in the decomposition 
(\ref{recombI3}).} 
solution $b$ reads $b = b_{0} +  {\cal K}er \, ({\cal S}) 
=  b_{0} + x \, v_{(3)}|_{\sigma_{3} = 0}$ with $x$ arbitrary scalar and
$b_{0} \equiv {\cal T}_{0} \cdot e$, leading to 
$B_{0} = e^{T} \cdot {\cal T}_{0} \cdot e$. 
The {\tt GOLEM} formalism thus applies also, using $b =b_{0}$ and $B=B_{0}$, 
when standing precisely at the peculiar configurations.
Yet slightly away from these peculiar configurations 
the {\tt GOLEM} ingredients defined by $b= {\cal S}^{-1} \cdot e$ 
separately show {\em discontinuities}\footnote{More precisely $b_{0}$ is
equal to the directional limit $s_{2} \to 0$ of $b$ along the direction 
$s_{-} = 0$ i.e. $t = \infty$. The discontinuities are meant 
for all other, finite $t$ directions.} w.r.t. those given by 
$b=b_{0}$ precisely at the peculiar configurations; this discontinuity comes
from the contribution to $b$ coming from the (divergent) component along
$v_{(3)}$, which have no counterpart in $b_{0}$. Notwithstanding, these 
individual discontinuities are artefacts in the sense that they cancel out in 
the reduction formula when put altogether, as discussed above.

%% file: concl.tex
\mysection{Summary and outlook}\label{concl}

In this article, we provided a representation of one-loop, 3-point functions
in 4 and 6 dimensions in the form of one dimensional representations in the
general case with complex masses. These one-dimensional integral representation 
have the virtue to avoid the appearance of factitious negative powers of Gram 
determinants, and are therefore numerically stable and remain rather/relatively
fast to compute numerically. We addressed the two cases at hand separately: 
the generic case when $\detg $ becomes small whereas $\dets$ remains finite, 
and the trickier specific case when both $\detg $ and $\dets$ become 
concomitantly small. Here we presented the ``existence proof" for scalar 
integrals, but the method applies to tensor integrals as well, i.e. loop 
integrals involving integer powers of Feynman parameters in the denominators of 
their integrands. 

\vspace{0.3cm}

\noindent
A forthcoming article will continue the present one by the 
similar treatment of one-loop 4-point functions. The latter proved to be quite 
more involved than the 3-point case, we thus preferred to split it from the
present article. These two will be supplemented by a dedicated treatment of the 
specific mixed case involving both finite (complex) masses, and some zero masses
triggering infrared issues. In the meantime we also found an alternative
approach leading to a derivation of integral representation which 
is perhaps simpler and also makes the algebraic nature of the ingredients 
involved more transparently related to the {\tt GOLEM} reduction algorithm 
both for 3-point and 4-point functions, this approach will be presented in 
a separate article. Last, this approach will be fully implemented in the next
version of the {\tt GOLEM95} library in Fortran 95. We will provide various
numerical tests of numerical stability at this occasion.

%% file: appendix-magicidientities.tex

\mysection{Useful algebraic identities among determinants}
\label{algebraic-identities}

Equation (\ref{eqI41*}) used in sect. \ref {sect3} relates various ingredients 
of the reduction formula involving the one loop three point function. Similar 
identities can be found and used  in the case of four point functions.
These properties can be traced back to general algebraic 
identities between the determinant of a square matrix and minors of this 
matrix, referred to as ``Jacobi identities for determinant ratios" 
\cite{encyclopedymath,jacobi-id}. 
This appendix reminds these general identities and 
specifies them to the case useful for the present work. Beforehand we remind a
few properties useful in this respect.

\subsection{Preliminaries}

Let us first recall a few useful properties which we state in general for
arbitrary $N$ not just $N=3$. Consider the kinematic $N \times N$ matrix 
$\cals$ associated with a 
given one-loop $N$-point diagram generalizing eq. (\ref{eqcals}) for any $N$. 
We single out the line and column $a$, and consider the corresponding 
$(N-1) \times (N-1)$ 
Gram matrix $G^{(a)}$ associated to $\cals$, generalizing (\ref{eqGRAMMAT}) and
the $(N-1)$-column vector $V^{(a)}_{i}$ generalizing eq. (\ref{eqVJA}).
Let us choose $a = N$ to fix the ideas and make formulas simpler; the results 
obtained can be straightforwardly generalized to any $a$. 

\vspace{0.3cm}

\noindent
1. Let us subtract the last column of $\cals$ from every column $j$, then 
subtract the last line from every line $i$ in the intermediate
matrix thus obtained, with $1 \leq i,j \leq N-1$. This defines the $N \times N$ 
matrix $\widehat{\cals^{(N)}}$ given by:
\begin{eqnarray}
\widehat{\cals^{(N)}}
& = & 
\left[ 
 \begin{array}{rcc}
 - \; G^{(N)}_{ij} & |  & V^{(N)}_{i} \\
 ---           & +  & ---        \\
V^{^(N)}_{j}       & |  & S_{NN}
 \end{array}
\right]
\label{e3}
\end{eqnarray}
generalizing eqs. (\ref{eqGRAMMAT}) and (\ref{eqVJA}). 
The determinants $\dets$ and $\det (\widehat{\cals^{(N)}})$ are 
equal. $\det (\widehat{\cals^{(N)}})$ is Laplace-expanded 
according to its $N^{th}$ line as:
\begin{eqnarray}
\det  (\widehat{\cals^{(N)}})
& = &
S_{NN} \, \det \! \left( - \, G^{(N)} \right) \, + \, 
\sum_{j=1}^{N-1} 
(-1)^{N+j} \, V^{(N)}_{j} \, 
\det \! \left( \widehat{\cals^{(N)}}^{ \{N\}}_{ \{j\}} \right)
\label{e5}
\end{eqnarray}
where $\widehat{\cals^{(N)}}^{ \{N\}}_{ \{j\}}$ is the $(N-1) \times (N-1)$ 
matrix obtained from 
$\widehat{\cals^{(N)}}$ by suppressing its line $N$ and column $j$. 
Using this notation we have in particular:
\begin{equation}\label{e5b}
- \, G^{(N)} =  \widehat{\cals^{(N)}}^{ \{N\}}_{ \{N\}}
\end{equation}
The determinant $\det \! 
\left(\widehat{\cals^{(N)}}^{ \{N\}}_{ \{j\}} \right)$ may 
in turn be Laplace-expanded with respect to its last column:
\begin{eqnarray}
\det \! \left( \widehat{\cals^{(N)}}^{ \{N\}}_{ \{j\}} \right)
& = &
\sum_{i=1}^{N-1}
(-1)^{i+(N-1)} V^{(N)}_{i} \, 
(-1)^{N-2} \, \left\{ (-1)^{i+j} \, 
\left( \widetilde{G^{(N)}} \right) _{ij} \right\}
\label{e6}
\end{eqnarray}
In the r.h.s. of eq. (\ref{e6}), 
the factor $(-1)^{N-2}$ comes from the explicit minus sign 
in $- \, G^{(N)}$; $(-1)^{i+j} \, (\widetilde{G^{(N)}})_{ij}$ is the minor of 
the Gram matrix element $G^{(N)}_{ij}$. Substituting into eq. (\ref{e5}) we get:
\begin{eqnarray}
\dets
& = &
(-1)^{N-1} 
\left[
 \cals_{NN} \, \det \! \left( G^{(N)} \right) \, + \, 
 V^{(N) \, T} \cdot \widetilde{G^{(N)}} \cdot V^{(N)}
\right]
\label{e7}
\end{eqnarray}
hence eq. (\ref{detsdetggtilde}).

\vspace{0.3cm}

\noindent
2. The coefficients $b_{i}$ 
in the {\tt GOLEM} $N$-point reduction algorithm are defined by
\begin{equation}\label{e8}
\sum_{j=1}^{N} \cals_{ij} b_{j} = 1, \;\;\;\; i=1, \cdots, N
\end{equation}
Singling out $b_{N}$ in eq. (\ref{e8}) corresponding to $i=N$, and
subtracting eq. (\ref{e8}) for $i=N$ from eq. (\ref{e8}) for every 
$i=1,\cdots,N-1$, eq. (\ref{e8}) may alternatively be 
rewritten in terms of $G^{(N)}$ and $V^{(N)}_{i}$ as:
\begin{eqnarray}
\sum_{j=1}^{N-1} b_{j} \, + \, b_{N}
& = & 
B
\label{e10a}\\
\sum_{j=1}^{N-1} G^{(N)}_{ij} \, b_{j} 
& = &
B \, V^{(N)}_{i}, \;\;\;\; i=1,\cdots, N-1
\label{e10b}\\
\sum_{j=1}^{N-1} V^{(N)}_{j} \, b_{j} 
& = & 
1 - B \, \cals_{NN}
\label{e10c}
\end{eqnarray}
When $G^{(N)}$ is invertible, eq. (\ref{e10b}) is solved as:
\begin{eqnarray}
b_{j} 
& = &
B \, \sum_{k=1}^{N-1} \left[ G^{(N)} \right]^{-1}_{jk} V^{(N)}_{k}
=
B \, \left[ \det \! \left( G^{(N)} \right) \right] ^{-1} \, 
\sum_{k=1}^{N-1} \left( \widetilde{G^{(N)}} \right) _{jk} V^{(N)}_{k}
\label{e11}
\end{eqnarray}
where the matrix $\widetilde{G^{(N)}}$ is the matrix of cofactors. 
Thus, using eq. (\ref{e7}): 
\begin{eqnarray}
\sum_{j=1}^{N-1} V^{(N)}_{j} b_{j}
& = & 
B \, \left[ \det \! \left( G^{(N)} \right) \right] ^{-1} \,
V^{(N) \, T} \cdot \widetilde{G^{(N)}} \cdot V^{(N)}
\nonumber\\
& = &
B \, (-1)^{N-1} \frac{\dets}{\det (G^{(N)})} \, - B \, \cals_{NN}
\label{e12}
\end{eqnarray}
Comparing eqs. (\ref{e12}) and (\ref{e10c}) yields:
\begin{equation}\label{e13}
B = (-1)^{N-1} \frac{\det (G^{(N)})}{\dets}
\end{equation}
and $b_{N}$ is obtained by solving eq. (\ref{e10a}). Introducing
\begin{equation}\label{e14}
\overline{b}_{j} \equiv \, b_{j} \, \dets 
\end{equation}
and, using eq. (\ref{e6}), eq. (\ref{e11}) reads:
\begin{eqnarray}\label{e15}
\overline{b}_{j} 
& = & 
(-1)^{N-1} \, \left[ \widetilde{G^{(N)}} \cdot V^{(N)} \right]_{j}
\label{e15a}\\
& = & 
(-1)^{j+N-2} \, 
\det \! 
\left( \widehat{\cals^{(N)}}^{ \{N\}}_{ \{j\}} \right),
\;\;\;\; j=1, \cdots, N-1
\label{e15b}
\end{eqnarray}

\subsection{The identity (\ref{eqI41*}) and 
Jacobi identities for determinants ratios}

As shown below, identity  (\ref{eqI41*}) is a special case of the following 
general property \cite{encyclopedymath}. Let $A$ be any $n \times n$ matrix, and
$A^{ \{i_{1}, \cdots, i_{r}\}}_{ \{k_{1}, \cdots, k_{r}\}}$ the matrix obtained 
from $A$ by suppressing the lines $i_{1}, \cdots, i_{r}$ and 
columns $k_{1}, \cdots, k_{r}$.
Then, for any $i_{1} < i_{2}$ and  $k_{1} < k_{2}$:
\begin{equation}
\det \! \left( \, A \, \right) \, 
\det \! \left( A^{ \{i_{1}, i{2}\}}_{ \{k_{1}, k_{2}\}} \right)
=
\det \! \left( A^{ \{i_{1}\}}_{ \{k_{1}\}} \right) \,
\det \! \left( A^{ \{i_{2}\}}_{ \{k_{2}\}} \right)
\, - \, 
\det \! \left( A^{ \{i_{1}\}}_{ \{k_{2}\}} \right)\,
\det \! \left( A^{ \{i_{2}\}}_{ \{k_{1}\}} \right)
\label{e19}
\end{equation}
Indeed, let us specify $A = \widehat{\cals^{(N)}}$ in the identity (\ref{e19})
and give the explicit forms of the other quantities obtained by suppressing appropriate 
lines and columns.
Let us take any $i \neq N$. 
\vspace{0.3cm}

\noindent
The $(N-2) \times (N-2)$ matrix $\widehat{\cals^{(N)}}^{ \{i,N\}}_{ \{i,N\}}$ 
is nothing but the matrix $- G^{i}$, thus
\begin{eqnarray}
\det \! \left( G^{\{i\}} \right) 
\det \! \left( \widehat{\cals^{(N)}}^{ \{i,N\}}_{ \{i,N\}} \right)
& = & 
(-1)^{N-2} \, \det \! \left( G^{\{i\}} \right) 
\label{e22}
\end{eqnarray}
Furthermore, , we first notice that, for $A$  symmetric, 
\begin{equation}\label{e20a}
A^{ \{i\}}_{ \{k\}} 
=
\left( 
A^{ \{k\}}_{ \{i\}} 
\right)^{T}
\end{equation}
thus 
\begin{equation}\label{e20}
\det \! \left( 
A^{ \{i\}}_{ \{k\}} 
\right) 
=
\det \!
\left( 
A^{ \{k\}}_{ \{i\}} 
\right)
\end{equation}from eqs. (\ref{e15b}) and (\ref{e20}), we have:
\begin{eqnarray}
\left( \overline{b}_{i} \right) ^{2}
& = &
\left( 
 \det \! \left( \widehat{\cals^{(N)}}^{ \{N\}}_{ \{i\}} \right)
\right) ^{2} 
= 
\det \! \left( \widehat{\cals^{(N)}}^{ \{N\}}_{ \{i\}} \right)
\,
\det \! \left( \widehat{\cals^{(N)}}^{ \{i\}}_{ \{N\}} \right)
\label{e21}
\end{eqnarray}
Besides,
\begin{equation}\label{e23}
\det \! \left( \widehat{\cals^{(N)}}^{ \{N\}}_{ \{N\}} \right)
= 
(-1)^{N-1} \,\det \! \left( G \right) 
\end{equation}
In the case at hand, identity (\ref{e19}) thus reads:
\begin{eqnarray}
\lefteqn{
 \underbrace{
 \det \! \left( \widehat{\cals^{(N)}} \right)
 }_{\det ({\cal S})} \, 
 \underbrace{
 \det 
 \left( 
  \widehat{\cals^{(N)}}^{ \{i,N\}}_{ \{i,N\}} 
 \right)
 }_{(-1)^{N-2}\, \det \left( G^{\{i\}} \right)}
}
\nonumber\\
& = &
\underbrace{
\det \!
\left( 
 \widehat{\cals^{(N)}}^{ \{i\}}_{ \{i\}} 
\right)
}_{\det \left( {\cal S}^{ \{i\}} \right) } \,
\underbrace{
\det \! 
\left( 
\widehat{\cals^{(N)}}^{ \{N\}}_{ \{N\}} 
\right)
}_{(-1)^{N-1} \, \det \left( G \right) }
\, - \, 
\underbrace{
\det \!
\left(
 \widehat{\cals^{(N)}}^{ \{i\}}_{ \{N\}} 
\right) \,
\det \!
\left( 
 \widehat{\cals^{(N)}}^{ \{N\}}_{ \{i\}} 
\right)
}_{\left( \overline{b}_{i} \right) ^{2}}
\label{e24}
\end{eqnarray}
i.e.
\begin{eqnarray}
\overline{b}^{2}_{i} 
& = &
(-1)^{N-1} \, 
\left[
 \det \! \left( \, \cals \, \right) \, \left( G^{\{i\}} \right) 
 \, + \,
 \det \! \left( \cals^{ \{i\}} \right) \, \det \! \left( G \right) 
\right]
\label{e24b}
\end{eqnarray}
Specifying $N=3$ in the present case of interest gives eq. (\ref{eqI41*}), with 
\begin{eqnarray}
\gamma^{\prime \prime}_{j} = \frac{1}{2} \, \det \! \left( G^{\{j\}} \right) 
& \, &
\Delta_{j} = - \, \det \! \left( \cals^{\{j\}} \right)
\label{emachin}
\end{eqnarray}
and where $G^{\{j\}}$ is the $(N-2) \times (N-2)$ Gram matrix associated to 
$\cals^{\{j\}}$ and obtained from it via a procedure similar to the one leading 
to eq. (\ref{e3}). q.e.d.

%% file: appendix-detg-zero-kinematics.tex
\mysection{Kinematics leading to a vanishing $\detg$}
\label{detg-zero-kinematics}\label{appendix-a2}

This appendix supplements the discussion on the kinematics leading to a
vanishing $\detg$ provided in subsection \ref{sss221}. 

\subsection{General considerations}

Let us consider a set $\{p_{i}, i = 1 \cdots , N-1\}$ of $N-1$ four-momenta in 
Minkowski space, their Gram matrix\footnote{The overall factor 2 in the 
definition of $G$ is actually irrelevant in the present discussion, we keep it 
only for notation consistency with the bulk of the article.}
$G_{ij} = 2 \, (p_{i} \cdot p_{j})$, and the linear system given by
\begin{eqnarray}
\sum_{j=1}^{N-1} G_{ij} \, x_{j} = 0 & , & i=1, \cdots, N-1
\label{systgram}
\end{eqnarray}
A vanishing $\detg$ means the existence of a set of scalars
$\{x_{j}, j=1,\cdots, N-1\}$ not all vanishing and solution of the system 
(\ref{systgram}). Multiplying eq. (\ref{systgram}) for each $i=1, \cdots,N-1$
by $x_{i}$ and summing over $i$ leads to the condition
\begin{eqnarray}
l^{2} = 0 
& , & 
l \equiv \sum_{j=1}^{N-1} x_{j} \, p_{j} 
\label{def-l}
\end{eqnarray}
which means that (i) $l$ vanishes i.e. the $\{p_{i}\}$ are 
linearly dependent momenta, or that (ii)  $l$ is lightlike and eq.
(\ref{systgram}) is the orthogonality condition $(l \cdot p_{i}) = 0$, 
$i=1, \cdots, N-1$ \cite{kinem}.
Let us focus on case (ii) assuming furthermore the $\{p_{j}\}$ to be
linearly independent\footnote{If both properties (i) and (ii) are
simultaneously fulfilled, then the rank of the $(N-1) \times (N-1)$ matrix
$G$ is (at most) $(N-3)$, corresponding to quite degenerate
configurations. For example for $N=3$ this corresponds to all four-momenta 
lightlike and collinear to each other, for which $G$ identically vanishes. 
For $N=4$ it corresponds to two spacelike and two collinear lightlike external 
momenta being a linear combination of the two spacelike ones. 
None of these cases are involved in NLO calculation of processes relevant 
e.g. for collider physics.}. The orthogonality condition requires that none 
of the $p_{j}$ be timelike, and $p_{N} \equiv - \sum_{j=1}^{N-1} p_{i}$ is 
orthogonal to $l$ too, thus cannot be timelike either.

\vspace{0.3cm}

\noindent
If one of the $p_{j}$, say $p_{1}$, is lightlike, $l$ is proportional to 
$p_{1}$, and all the $p_{j\neq 1}$ shall be spacelike. Were $p_{N}$ lightlike 
it should be $\propto p_{1}$, which would contradict the extra assumption of 
linear independence of
the $\{p_{i}\}$, $i = 1 \cdots , N-1$, thus $p_{N}$ shall be spacelike too.
These requests impose a steric constraint on $N$ w.r.t the 
spacetime dimension $d=4$. As seen by decomposing $p_{j}$ as 
$(p^{0}_{j}/p_{1}^{0}) \, p_{1} + q_{j}$ for $j = 2, \cdots, N-1$ in a frame
chosen such that $p_{1} = p_{1}^{0}(1; \vec{0}_{\perp}; \epsilon,)$, with 
$\epsilon = \pm$ and $q_{j} = (0; \vec{q}_{\perp \, j}; 0)$, the maximal number 
of possibly independent $q_{j}$ is $d-2=2$ i.e. $N$ shall be $\leq 4$.
Besides, for $N \geq 4$, NLO calculations involve no one-loop $N$-point 
function with external momenta all spacelike but one lightlike, neither in 
collision nor in decay processes: alternative (ii) only occurs for $N=3$ for any
NLO purpose.

\vspace{0.3cm}

\noindent
If none of the $p_{j}$ $j =1,\cdots, N-1$ is lightlike, all of them shall be 
spacelike. The momentum $p_{N}$ shall be either lightlike - hence 
proportional to $l$: one is driven back to the previous case; see the $N=3$ 
case below - or spacelike. The latter case is submitted to a 
similar steric constraint as above, as seen by trading one of the $p_{j}$ for 
$l$; no such configuration matters at NLO whatever $N$.

\vspace{0.3cm}

\noindent
In summary, for any practical purpose at NLO, a vanishing $\detg$ can happen 
for a linearly independent kinematic configuration only in the case $N=3$.
Otherwise the configurations with vanishing $\detg$ correspond to linearly
dependent four-momenta.  

\subsection{Focus on $N=3$}

This appendix elaborates on the case $N=3$ involved in subsubsec. \ref{sss221}, with 
$p_{1}$ and $p_{3}$ linearly independent and spacelike.
We parametrize the lightlike combination $l$ (defined up to an overall 
multiplicative constant) as $l = p_{1} - x \, p_{3}$.
The orthogonality conditions (implying that $l$ is lightlike) read:
\begin{eqnarray}
( l \cdot p_{1}) & = & s_{1} - x \, (p_{1} \cdot p_{3})
\label{cond1}\\
( l \cdot p_{3}) & = & (p_{1} \cdot p_{3}) - x \,  s_{3}
\label{cond2}
\end{eqnarray} 
The vanishing $\detg = s_{1} \, s_{3} -(p_{1} \cdot p_{3})^{2}$ ensures the
compatibility of eqs. (\ref{cond1}) and (\ref{cond2}) in $x$ and
\begin{eqnarray}
x = \frac{(p_{1} \cdot p_{3})}{s_{3}} =  
- \, \mbox{sign}(p_{1} \cdot p_{3}) \, \sqrt{\frac{s_{1}}{s_{3}}}
\label{paramlightlike2}
\end{eqnarray}
The condition $\detg = 0$ also implies that 
$s_{2} = s_{1} + 2 \, (p_{1} \cdot p_{3}) + s_{3}$ can be written
\begin{eqnarray}
s_{2} 
& = & 
- \, 
\left( 
 \sqrt{-s_{1}} - \, \mbox{sign}(p_{1} \cdot p_{3}) \, \sqrt{-s_{3}}
\right)^{2} 
\leq 0 
\label{s2}
\end{eqnarray}
Therefore $s_{2} = 0$ iff $\mbox{sign}(p_{1} \cdot p_{3}) = +$ and 
$s_{1} = s_{3}$, in which case $x = - \, 1$ and $p_{2} = - \, l$. 
Otherwise $s_{2} < 0$.

%% file: appendix-spectral-decomp-S.tex
\mysection{Spectral features of ${\cal S}$ for $N=3$
}\label{spectral-decomp-S}

This appendix gathers the spectral properties of ${\cal S}$ for $N=3$ which 
are further used in Appendix \ref{calcbj-B}. 

\vspace{0.3cm}

\noindent
Accounting for the condition $m_{1}^{2} = m_{2}^{2} \equiv m^{2}$ and the
parametrization used in subsubsec. 2.2.2, the kinematic matrix ${\cal S}$ reads:
\begin{eqnarray}
{\cal S} 
=
\left[
 \begin{array}{ccc}
- \, 2 m^{2} & s_{2} - 2 \, m^{2} & s_{+}+s_{-} -(m^{2}+m^{2}_{3})\\
s_{2} - 2 \, m^{2} & - \, 2 m^{2} & s_{+}-s_{-} -(m^{2}+m^{2}_{3})\\ 
s_{+}+s_{-} -(m^{2}+m^{2}_{3}) & s_{+}-s_{-} -(m^{2}+m^{2}_{3}) &
- 2 \, m^{2}_{3}
 \end{array}
\right]
&& \;\;\;
\label{a31}
\end{eqnarray} 
Let us compute the eigenvalues $\sigma_{1,2,3}$ of ${\cal S}$ and
the corresponding eigenvectors $v_{(1,2,3)}$ in the regime $\detg \to 0$,  
$\dets \to 0$ corresponding to $s_{-} \to 0$, $s_{2} \to 0$. Since 
$\sigma_{3} \to 0$ whereas $\sigma_{1,2}$ remain nonvanishing in this regime, in
order to correctly get the coefficients $b_{j}$ and $B$ in eqs. (\ref{bj-1}) 
and (\ref{B-1}) respectively in subsubsec. \ref{ss222}, we shall keep the
leading  dependence on $s_{-}, s_{2}$ in $\sigma_{3}$ and in the components of
the  corresponding normalized eigenvector $v_{(3)}$, whereas $s_{-}$ and $s_{2}$
can  be safely put to zero to first approximation in $\sigma_{1,2}$ and the 
corresponding normalized eigenvectors $v_{(1,2)}$. This is the approximation to 
which we provide the algebraic results below.

\subsection{Eigenvalues}\label{sect-a31}

The characteristic polynomial ${\cal P}_{\cal S}(s)$ of ${\cal S}$ is:
\begin{eqnarray}
{\cal P}_{\cal S}(s) 
& \equiv & 
\det \left( {\cal S} - s \, 1\!\mbox{I}_{3} \right)
\nonumber\\
& = & 
- \, 
\left\{ 
 s^{3} - \left( \mbox{tr} ({\cal S}) \right) \, s^{2} + 
 \frac{1}{2} \, 
 \left[ 
  \left( \mbox{tr} ({\cal S}) \right)^{2} - 
  \mbox{tr} \left( {\cal S}^{2} \right)
 \right] \, s - \dets
\right\}
 \label{a32}
\end{eqnarray}
{\bf The small eigenvalue $\sigma_{3}$}

\vspace{0.3cm}

\noindent
The eigenvalue $\sigma_{3}$ vanishing as $\dets$ 
may be extracted from eq. (\ref{a32}) rewritten 
\begin{equation}
\sigma_{3} 
=
\frac{2 \, \dets}
{\left(  \mbox{tr} ({\cal S}) \right)^{2} - 
 \mbox{tr} \left( {\cal S}^{2} \right)} 
+
\frac{2 \; \mbox{tr} ({\cal S})}
{\left( \mbox{tr} ({\cal S}) \right)^{2} - 
 \mbox{tr} \left( {\cal S}^{2} \right)} \, 
\sigma_{3}^{2} - 
\frac{2}
{\left( \mbox{tr} ({\cal S}) \right)^{2} - 
 \mbox{tr} \left( {\cal S}^{2}\right)} \, 
\sigma_{3}^{3}
\label{a33}
\end{equation}
by an iteration generating an expansion in integer powers of 
$\dets$. The leading term of this expansion, itself truncated to
keep only the leading dependencies in $s_{2}$ and $s_{-}$, is given by:
\begin{eqnarray}
\sigma_{3} 
& = &
\frac{2 \, \dets|_{\mbox{\tiny trunc}}}
{\left[ \left( \mbox{tr} ({\cal S}) \right)^{2} - 
 \mbox{tr} \left( {\cal S}^{2}\right) \right]_{s_{-} = s_{2} = 0}} 
+ 
 \cdots
\label{a34}
\end{eqnarray}
Using
\begin{eqnarray}
\left( \mbox{tr} ({\cal S}) \right) ^{2}
& = &
2^{2} \, \left( 2 \, m^{2} \right) ^{2} + 
2^{2} \, \left( 2 \, m^{2} \right) \, \left( 2 \, m_{3}^{2} \right) +
\left( 2 \, m_{3}^{2} \right) ^{2}
\label{eq4b}\\
\mbox{tr} \left( {\cal S}^{2} \right) 
& = &
2^{2} 
\left\{ 
 \left[ s_{+} - \left( m^{2} + m_{3}^{2} \right) \right] ^{2} +
 (2 m^{2})^{2}
\right\} 
+ (2 m_{3}^{2})^{2}
\nonumber\\
& &
+ \, 4 \left( s_{-} \right)^{2} 
- \, (8 \, m^{2}) \, s_{2} + 2 \left( s_{2} \right)^{2}
\label{eq4e}
\end{eqnarray}
we have:
\begin{eqnarray}
\mbox{tr} \left( {\cal S} \right) ^{2} - 
\mbox{tr} \left( {\cal S}^{2} \right) 
& = &
- \, 4 \, \widetilde{\lambda} \, \left( 1 + \xi \right)
\label{eq4g1}\\
\xi 
& = &
\frac{1}{\widetilde{\lambda}} \, 
\left[ s_{-}^{2} - (2 \, m^{2}) \, s_{2} + \, \frac{1}{2} s_{2}^{2} \right]
\label{eq4g2}
\end{eqnarray}
whereas
\begin{equation}\label{eq4detS}
\dets = 2 \, \widetilde{\lambda}
\left\{ 
 \left[
 s_{2} + 
 \frac{4 m^{2}}{ \widetilde{\lambda}} \, s_{-}^{2} 
 \right]
 +
  s_{2} \,
 \left[ 
 \frac{m_{3}^{2}}{\widetilde{\lambda}} \, s_{2} -
 \frac{1}{ \widetilde{\lambda}} \, s_{-}^{2}
 \right]
\right\}
\end{equation}
We further truncate
\begin{eqnarray}
\left[ \mbox{tr} \left( {\cal S} \right) ^{2} -
 \mbox{tr} \left( {\cal S}^{2} \right) \right]_{s_{-} = s_{2} = 0}
& = &
- \, 4 \, \widetilde{\lambda} 
\label{eq4e}
\\
\dets|_{\mbox{\tiny trunc}} 
& = &
2 \, \widetilde{\lambda}
\left( s_{2} + \frac{4 m^{2}}{ \widetilde{\lambda}} \, s_{-}^{2} \right) + \cdots
\end{eqnarray}
The eigenvalue $\sigma_{3}$ thus has the following approximate expression:
\begin{eqnarray}
\sigma_{3}
& = &
- \; 
\left(  s_{2}  + \frac{4 m^{2}}{\widetilde{\lambda}} \, s_{-}^{2} \right) 
+ \cdots
\label{eq5}
\end{eqnarray}
in which the terms dropped are of order $s_{2}^{2}$, $s_{2} \, s_{-}^{2}$, 
$s_{-}^{4}$ and higher. 

\vspace{0.3cm}

\noindent
{\bf The two non vanishing eigenvalues $\sigma_{1,2}$}

\vspace{0.3cm}

\noindent
The two other eigenvalues $\sigma_{1,2}$ are obtained from the 
factorization of ${\cal P}_{\cal S}(s)$ as:
\begin{eqnarray}
{\cal P}_{\cal S}(s)
& = &
- \, \left( s - \sigma_{3} \right) \, \left( s^{2} - {\cal A} \, s + \, {\cal B}
\right) 
\label{eq6}
\end{eqnarray}
which requires
\begin{eqnarray}
{\cal A} + \sigma_{3} 
& = & 
\mbox{tr} ({\cal S})
\label{eq7a}\\
{\cal B} + \sigma_{3} \, {\cal A} 
& = &
\frac{1}{2} 
\left[
 \left( \mbox{tr} ({\cal S}) \right) ^{2} - 
 \mbox{tr} \left( {\cal S} ^{2} \right)
\right]
\label{eq7b}\\
\sigma_{3} \, {\cal B} 
& = & 
\dets
\label{eq7c}
\end{eqnarray}
The approximation corresponding to $s_{-} = s_{2} = 0$ 
in eqs. (\ref{eq7a}) - (\ref{eq7c}) replaces
\begin{eqnarray}
{\cal A} 
& \to & 
{\cal A}_{_{\emptyset}}
=  
\mbox{tr} ({\cal S}) = - \, 2 \left( 2 \, m^{2} + m_{3}^{2} \right)
\label{eq8a}\\
{\cal B} 
& \to & 
{\cal B}_{_{\emptyset}} 
=
\frac{1}{2} \, 
\left[
 \left( \mbox{tr} ({\cal S}) \right) ^{2} - 
 \mbox{tr} \left( {\cal S} ^{2} \right)
\right]_{s_{-} = s_{2} = 0} 
= 
- \, 2\, \widetilde{\lambda} 
\label{eq8c}
\end{eqnarray}
and the ``zeroth order" approximations of $\sigma_{1,2 }$ are given by:  
\begin{eqnarray}
\sigma_{^{1}_{2} \, \emptyset} 
& = &
- \, \left( 2 \, m^{2} + m_{3}^{2} \right) \pm 
\sqrt{\left( 2 \, m^{2} + m_{3}^{2} \right) ^{2} + 2 \, \widetilde{\lambda}}
\label{eq9}
\end{eqnarray}

\subsection{Eigenvectors}\label{sect222}

{\bf The eigenvector $v_{3}$  associated with $\sigma_{3}$}

\vspace{0.3cm}

\noindent
The components $x,y,z$ of $v_{3}$ are solutions of
the degenerate system:
\begin{eqnarray}
- \left( \sigma_{3}  + 2 \, m^{2} \right) \, x 
+ \left( s_{2} - 2 \, m^{2} \right) \, y 
+ \left( s_{+} + s_{-} -(m^{2}+m_{3}^{2}) \right) \, z 
=0&& \;\;\;\;
\label{eq10a}\\
  \left( s_{2} - 2 \, m^{2} \right) \, x  
- \left( \sigma_{3}  + 2 \, m^{2} \right) \, y  
+\left(  s_{+} - s_{-} - (m^{2}+m_{3}^{2}) \right) \, z
=0 && \;\;\;\;
\label{eq10b}\\ 
  \left( s_{+}+s_{-} -(m^{2}+m_{3}^{2}) \right) \, x 
+ \left( s_{+}-s_{-} -(m^{2}+m_{3}^{2}) \right) \, y
- \left( \sigma_{3} + 2 \, m_{3}^{2} \right) \, z 
=0 &&\;\;\;\;
\label{eq10c}
\end{eqnarray}
Subtracting (\ref{eq10a}) from eq. (\ref{eq10b}) yields:
\begin{eqnarray}
\left( s_{2} + \sigma_{3} \right) \, \left( x - y \right) - 2 \, s_{-} \, z
& = & 0
\label{eq10d}
\end{eqnarray}
Since 
\[
s_{2} + \sigma_{3} 
\sim 
- \; \frac{(4 m^{2})}{\widetilde{\lambda}} \, s_{-}^{2}
\]
eq. (\ref{eq10d}) tells that $z = {\cal O}(s_{-} (x-y))$: $x$ and $y$ being
bounded, $z$ thus vanishes at least as ${\cal O}(s_{-})$ in the limit 
$s_{-} \to 0$. We shall keep the leading dependence on $s_{-}$ and $s_{2}$ in 
the components of $v_{(3)}$.

\vspace{0.3cm}

\noindent
Up to an overall normalization, $x$, $y$ and $z$ are given by: 
\begin{eqnarray}
x 
& = &
- \, \widetilde{\lambda} \, + 2\, (m^{2}+m_{3}^{2}) \, \sigma_{3}\, 
+ \, 2 \, \left( s_{+} -(m^{2}+m_{3}^{2}) \right) s_{-} 
- s_{-} ^{2} 
+ \cdots
\label{eq11a}\\
y 
& = &
\;\;
\widetilde{\lambda} 
- 2 \, (m^{2}+m_{3}) \, \sigma_{3} \,
- \left[ 1 + \frac{8 \, m^{2} \, m_{3}^{2}}{\widetilde{\lambda}} \right] \,
  s_{-}^{2}
+  \cdots
\label{eq11b}\\
z 
& = &
\left( 4 \, m^{2} \right) \, s_{-} -
\left( s_{+} - (m^{2}+m_{3}^{2}) \right) \, 
\frac{4 \, m^{2}}{\widetilde{\lambda}} \, s_{-}^{2}
+  \cdots
\label{eq11c}
\end{eqnarray}
In eqs. (\ref{eq11a})-(\ref{eq11c}), the dependence on $s_{2}$ has been
traded for $s_{-}$ and $\sigma_{3}$ up to terms neglected at the
approximation retained.
Introducing\footnote{In what follows it is  not necessary to normalize the 
vector $l_{(3)}$ to 1.}
\begin{eqnarray}
v_{(3) \, \emptyset}
=
\frac{1}{\sqrt{2}} \, 
\left[
 \begin{array}{r}
    1 \\
  - \, 1 \\
    0 
 \end{array}
\right]
& , &
l_{(3)} 
=
\left[
 \begin{array}{c}
    \left( s_{+} - (m^{2}+m_{3}^{2}) \right) \\
    \left( s_{+} - (m^{2}+m_{3}^{2}) \right) \\
    4 \, m^{2} 
 \end{array}
\right]
\label{eq-defl3}
\end{eqnarray}
the unnormalized eigenvector $v_{(3)}^{unnorm}$ given by eqs. 
(\ref{eq11a})-(\ref{eq11c}) can be written:
\begin{eqnarray}
v_{(3)}^{unnorm}
& = &
- \, \widetilde{\lambda} \sqrt{2} \, 
\left( 
 1 - \frac{2 \, ( m^{2} + m_{3}^{2})}{\widetilde{\lambda}} \, \sigma_{3} 
 - \frac{\left( s_{+} - (m^{2}+m_{3}^{2}) \right)}{\widetilde{\lambda}} \, 
 s_{-} - \frac{4 \, m^{2} \, m_{3}^{2}}{\widetilde{\lambda}^{2}} \, s_{-}^{2} 
\right) 
v_{(3) \, \emptyset}
\nonumber\\
&&
+ \,  s_{-} 
\left( 
 1 - \frac{\left( s_{+} - (m^{2}+m_{3}^{2}) \right)}{\widetilde{\lambda}} \, 
 s_{-} 
\right) \, l_{(3)}  
\, +  \cdots
\label{eq12bis} 
\end{eqnarray}
The vector $v_{(3) \, \emptyset}$ is the normalized eigenvector of ${\cal S}$
associated with the eigenvalue $\sigma_{3}=0$ when $\dets = 0$.
Let us notice that $(l_{(3)}^{T} \cdot v_{(3) \, \emptyset}) = 0$ and 
$(e^{T} \cdot v_{(3) \, \emptyset}) = 0$ where the vector $e$ was defined in 
eq. (\ref{def-e}) in subsubsec. \ref{ss222}. Once normalized by 
${\cal N}_{3}  \equiv - \, (v_{(3)}^{\mbox{\tiny unnorm} \, T} \cdot 
v_{(3)}^{\mbox{\tiny unnorm}}) ^{-1/2}$, the eigenvector $v_{(3)}$ is given by: 
\begin{eqnarray}
v_{(3)}
& = & 
\left(
 1 
 + {\cal O}(s_{-}^{2})
\right)
v_{(3) \, \emptyset} \,
- \, \frac{s_{-} }{\widetilde{\lambda}\sqrt{2}} \,  
\left( 1 + {\cal O}(s_{-}) \right) \, l_{(3)}  
\, + \cdots 
\label{eq14bis}
\end{eqnarray}
The ${\cal O}(s_{-}^{2})$ terms are no more explicited 
in eq. (\ref{eq14bis}) as they would contribute in Appendix \ref{calcbj-B} 
beyond the level of approximation retained only. The departure of $v_{(3)}$ 
from $v_{(3) \, \emptyset}$ in eq. (\ref{eq14bis}) does not depend explicitly 
on $\sigma_{3}$, it only depends on $s_{-}$.

\vspace{0.3cm}

\noindent
{\bf The eigenvectors $v_{^{(1)}_{(2)}}$ associated with $\sigma_{^{1}_{2}}$}

\vspace{0.3cm}

\noindent
The components $x_{^{1}_{2}},y_{^{1}_{2}},z_{^{1}_{2}}$ of the 
eigenvectors associated with $\sigma_{^{1}_{2}}$, are solutions of the degenerate 
system
\begin{eqnarray}
- ( \sigma_{^{1}_{2}}  + 2 \, m^{2} ) \, x_{^{1}_{2}} 
+ ( s_{2} - 2 \, m^{2} ) \, y_{^{1}_{2}} 
+ \left( s_{+} + s_{-} -(m^{2}+m_{3}^{2}) \right) \, z_{^{1}_{2}} 
=0 && \;\;\;\;
\label{eq15a}\\
  ( s_{2} - 2 \, m^{2} ) \, x_{^{1}_{2}}  
- ( \sigma_{^{1}_{2}}  + 2 \, m^{2}) \, y_{^{1}_{2}} 
+\left(  s_{+} - s_{-} - (m^{2}+m_{3}^{2}) \right) \, z_{^{1}_{2}}
= 0 && \;\;\;\;
\label{eq15b}\\ 
  \left( s_{+}+s_{-} -(m^{2}+m_{3}^{2}) \right) \, x_{^{1}_{2}} 
+ \left( s_{+}-s_{-} -(m^{2}+m_{3}^{2}) \right) \, y_{^{1}_{2}}
- ( \sigma_{^{1}_{2}} + 2 \, m_{3}^{2} ) \, z_{^{1}_{2}} 
=0 && \;\;\;\;
\label{eq15c}
\end{eqnarray}
Subtracting eq. (\ref{eq15a}) from eq. (\ref{eq15b}) yields:
\begin{eqnarray}
( s_{2} + \sigma_{^{1}_{2}} ) \, ( x_{^{1}_{2}} - y_{^{1}_{2}} ) - 
2 \, s_{-} \, z_{^{1}_{2}}
& = & 
0
\label{eq15d}
\end{eqnarray}
Since $|s_{2}| \ll |\sigma_{^{1}_{2}}| \neq 0$ and $z_{^{1}_{2}}$ remains bounded, 
$(x_{^{1}_{2}}-y_{^{1}_{2}})$ thus vanishes at least as ${\cal O}(s_{-})$ in the limit
$s_{-} \to 0$. 
In the zeroth order approximation corresponding to $s_{-} = s_{2} = 0$, 
$(x_{^{1}_{2}}-y_{^{1}_{2}})$ vanishes. Substituting this into eq. (\ref{eq15c}) 
the latter becomes:
\begin{eqnarray}
2\, \left( s_{+} -(m^{2}+m_{3}^{2}) \right) \, x 
- \left( \sigma_{^{1}_{2} \, \emptyset} + 2 \, m_{3}^{2} \right) \, z 
& = & 0 \;\;\;\;
\label{eq16}
\end{eqnarray}
which involves
\[
- \, ( \sigma_{^{1}_{2} \, \emptyset} + 2 \, m_{3}^{2}) = 
\left[ (2 \, m^{2} - m_{3}^{2}) \mp 
 \sqrt{(2 \, m^{2} + m_{3}^{2})^{2} + 2 \, \widetilde{\lambda}} \; 
\right]
\] 
Up to an overall normalization factor to be fixed below, $x_{^{1}_{2}}$, 
$y_{^{1}_{2}}$ and $z_{^{1}_{2}}$ are given by:
\begin{eqnarray}
x_{^{1}_{2} \, \emptyset} = y_{^{1}_{2} \, \emptyset} 
& = & 
\mp ( \sigma_{^{1}_{2} \, \emptyset} + 2 \, m_{3}^{2})
\label{eq17a}\\
z_{^{1}_{2} \, \emptyset} 
& = &
\mp \, 2 \, \left( s_{+} -(m^{2}+m_{3}^{2}) \right)
\label{eq17b}
\end{eqnarray}
The condition 
$(v^{T}_{(1) \, \emptyset} \cdot v_{(2) \, \emptyset}) = 0$ is fulfilled 
by eqs. (\ref{eq17a}), ( \ref{eq17b}) since:
\begin{eqnarray}
\lefteqn{x_{1 \, \emptyset} \, x_{2 \, \emptyset} + 
y_{1 \, \emptyset} \, y_{2 \, \emptyset} +
z_{1 \, \emptyset} \, z_{2 \, \emptyset}}
\nonumber\\ 
& = &
- \, 2 \, (\sigma_{1 \, \emptyset} + 2 \, m_{3}^{2})
(\sigma_{2 \, \emptyset} + 2 \, m_{3}^{2}) -
4 \, \left( s_{+} -(m^{2}+m_{3}^{2}) \right)^{2}
\nonumber\\
& = &
0
\label{eq17suite}
\end{eqnarray}
Identity (\ref{eq17suite}) will be used in Appendix \ref{calcbj-B}.
The normalization factor ${\cal N}_{^{1}_{2} \, \emptyset}$ required to 
normalize $|\!|v_{^{1}_{2} \, \emptyset}|\!|$ to $1$ is given by:
\begin{eqnarray}
{\cal N}_{^{1}_{2} \, \emptyset}
& = & 
\left[ 
 2 \, (\sigma_{^{1}_{2} \, \emptyset} + 2 \, m_{3}^{2})^{2} + 
 4 \, \left( s_{+} -(m^{2}+m_{3}^{2}) \right)^{2} 
\right] ^{-1/2}
\nonumber\\
& = & 
\left[ 
 \pm \, 2 \, \left( \sigma_{^{1}_{2} \, \emptyset} + 2 \, m_{3}^{2} \right) \, 
 \left( \sigma_{1 \, \emptyset} - \sigma_{2 \, \emptyset} \right) 
\right] ^{-1/2}
\label{eq17c}
\end{eqnarray}
Let us define the angle $\theta_{\emptyset}$ by
\begin{eqnarray}
\cos \theta_{\emptyset} 
& = & 
- \, \sqrt{2} \, ( \sigma_{1 \, \emptyset} + 2 \, m_{3}^{2}) \, 
{\cal N}_{1 \, \emptyset}
\label{eq18a}\\
\sin \theta_{\emptyset} 
& = & 
2 \, \left( s_{+} -(m^{2}+m_{3}^{2}) \right) \, {\cal N}_{1 \, \emptyset} 
\label{eq18b}
\end{eqnarray}
The normalized eigenvectors $v_{^{(1)}_{(2)} \, \emptyset}$ read:
\begin{eqnarray}
v_{(1) \, \emptyset}
=
\left[
 \begin{array}{r}
 \frac{1}{\sqrt{2}} \, \cos \theta_{\emptyset} \\
 \frac{1}{\sqrt{2}} \, \cos \theta_{\emptyset} \\
 - \, \sin \theta_{\emptyset}
 \end{array}
\right]
& , & 
v_{(2) \, \emptyset}
=
\left[
 \begin{array}{r}
 \frac{1}{\sqrt{2}} \, \sin \theta_{\emptyset} \\
 \frac{1}{\sqrt{2}} \, \sin \theta_{\emptyset} \\
 \cos \theta_{\emptyset}
 \end{array}
\right]
\label{eq18bb}
\end{eqnarray}
Together with $v_{(3) \, \emptyset}$ given by eq. (\ref{eq-defl3}) above
these eigenvectors define an orthonormal basis - namely the eigenbasis 
of ${\cal S}$ when $\dets = 0$. 
The overall signs have been chosen such that the orientation is direct i.e. 
$\det \, [v_{(1) \, \emptyset},v_{(2) \, \emptyset},v_{(3) \, \emptyset}] = + 1$.

%% file: appendix-calc-bj-B.tex
\mysection{Analysis of the reduction coefficients 
$(b_{0})_{j}$, $b_{j}$, $B_{0}$ and $B$ when 
$\detg$ and $\dets \to 0$}\label{calcbj-B}

This appendix provides a detailed analysis of the reduction coefficients
$(b_{0})_{j}$, $b_{j}$, $B_{0}$ and $B$ when $\detg, \dets \to 0$
providing the technical back-up to the discussion in subsubsecs. \ref{ss222} to 
\ref{ss224}. Introducing the vectors 
\begin{equation}\label{eq40}
n_{1} =
\left[
 \begin{array}{c}
 1 \\
 0 \\
 0
 \end{array}
\right]
\;\; 
, 
\;\;
n_{2} =
\left[
 \begin{array}{c}
 0 \\
 1 \\
 0
 \end{array}
\right]
\;\; 
, 
\;\;
n_{3} 
\left[
 \begin{array}{c}
 0 \\
 0 \\
 1
 \end{array}
\right]
\end{equation} 
the components of $b$ defined by eq. 
(\ref{bj-1}) can be expressed in the limit $\sigma_{3} \to 0$ in terms of those
of $b_{0} = {\cal T} \cdot e$ introduced in subsubsec. 
\ref{ss224} as:
\begin{eqnarray}
b_{j}(\sigma_{3} \to 0)
& = &
(b_{0})_{j} \, + \, 
\sigma_{3}^{-1} \, 
\left( e^{T} \cdot v_{(3)} \right)
\left( n_{j}^{T} \cdot v_{(3)} \right)
+ \cdots 
\label{eq39b}\\
(b_{0})_{j}
& = &
\sigma_{1 \, \emptyset}^{-1} \, 
\left( e^{T} \cdot v_{(1) \, \emptyset} \right) 
\left( n_{j}^{T} \cdot v_{(1) \, \emptyset} \right) + 
\sigma_{2 \, \emptyset}^{-1}  \, 
\left( e^{T} \cdot v_{(2) \, \emptyset} \right) 
\left( n_{j}^{T} \cdot v_{(2) \, \emptyset} \right)
\;\;\;\;
\label{eq39a}
\end{eqnarray}
where the column vector $e$ has been defined by eq. (\ref{def-e}), and where 
``$\cdots$" in eq. (\ref{eq39b}) stand for evanescent terms in the limit 
considered. As the saying goes, `a tedious but straightforward' algebraic 
juggling, sketched below, leads to the following expressions for the sought 
coefficients.

\vspace{0.3cm}

\noindent
{\bf (i) $(b_{0})_{3}$} 

\vspace{0.3cm}

\noindent
Using 
$\sigma_{1 \, \emptyset} \, \sigma_{2 \, \emptyset} = -2 \, \widetilde{\lambda}$,
we get:
\begin{eqnarray}
(b_{0})_{3}
& = &
- \, \frac{\sigma_{2 \, \emptyset}}{2 \, \widetilde{\lambda}}
\left( e^{T} \cdot v_{_{(1) \, \emptyset}} \right) 
\left( n_{3}^{T} \cdot v_{_{(1) \, \emptyset}} \right) 
- \, \frac{\sigma_{1 \, \emptyset}}{2 \, \widetilde{\lambda}}  
\left( e^{T} \cdot v_{_{(2) \, \emptyset}} \right) 
\left( n_{3}^{T} \cdot v_{_{(2) \, \emptyset}} \right) 
\;\;\;
\label{eq41}
\end{eqnarray}
This involves
\begin{eqnarray}
\left( e^{T} \cdot v_{_{(1) \, \emptyset}} \right)  
\left( n_{3}^{T} \cdot v_{_{(1) \, \emptyset}} \right)
& = &
- \, 
\left( 
 \sqrt{2} \, \cos \theta_{\emptyset} \, -  \, \sin \theta_{\emptyset}
\right) \, \sin \theta_{\emptyset}
\label{eq42a}\\
\left( e^{T} \cdot v_{_{(2) \, \emptyset}} \right) 
\left( n_{3}^{T} \cdot v_{_{(2) \, \emptyset}} \right)
& = &
\;\;\;\,\,
 \left( 
 \sqrt{2} \, \sin \theta_{\emptyset} \, + \, \cos \theta_{\emptyset}
\right) \, \cos \theta_{\emptyset}
\label{eq42b}
\end{eqnarray}
$(b_{0})_{3}$ takes the form:
\begin{eqnarray}
(b_{0})_{3}
& = &
- \, \frac{1}{2 \, \widetilde{\lambda}} \, 
\left\{
  \frac{1}{2} \,
  \left( \sigma_{1 \, \emptyset} + \sigma_{2 \, \emptyset} \right)
\right.
\nonumber\\
&&
\;\;\;\;\;\;\;\;  
\left.
  +
 \left[ 
  \frac{1}{2} \, 
  \left( \cos^{2} \theta_{\emptyset} - \sin^{2} \theta_{\emptyset} \right)
  + \sqrt{2} \, \sin \theta_{\emptyset} \, \cos \theta_{\emptyset}
 \right] \, \left( \sigma_{1 \, \emptyset} - \sigma_{2 \, \emptyset} \right)
\right\}
\;\;\;
\label{eq43}
\end{eqnarray}
With $\cos \theta_{\emptyset}$, $\sin \theta_{\emptyset}$ from eqs. 
(\ref{eq17c}) - (\ref{eq18b}) and using identity (\ref{eq17suite}), we have:
\begin{eqnarray}
\lefteqn{
\frac{1}{2} \, 
  \left( \cos^{2} \theta_{\emptyset} - \sin^{2} \theta_{\emptyset} \right)
  + \sqrt{2} \, \sin \theta_{\emptyset} \, \cos \theta_{\emptyset}
}
\nonumber\\
& = &
- \, \frac{1}
{\left( \sigma_{1 \, \emptyset} - \sigma_{2 \, \emptyset} \right)} \, 
\left[
 2 \, \left( s_{+} - (m^{2} + m_{3}^{2}) \right) + (2 \, m^{2} - m_{3}^{2})
\right]
\label{eq44b}
\end{eqnarray}
whereas $\sigma_{1 \, \emptyset} + \sigma_{2 \, \emptyset} 
= - 2(2 \, m^{2} + m_{3}^{2})$. Finally $(b_{0})_{3}$ reads:
\begin{eqnarray}
(b_{0})_{3}
& = &
\frac{1}{\widetilde{\lambda}} \,
\left[ \left( s_{+} - (m^{2} + m_{3}^{2}) \right) + (2 \, m^{2}) \right]
\nonumber\\
& = & 
\frac{1}{2 \, \widetilde{\lambda}} \, \left( e^{T} \cdot l_{(3)} \right)
\label{eq44b}
\end{eqnarray}
{\bf (ii) $b_{3}(\sigma_{3} \to 0)$} 

\vspace{0.3cm}

\noindent
The extra bit to be added to $(b_{0})_{3}$ to get 
$b_{3}(\sigma_{3} \to 0)$ is 
$\propto (e^{T} \cdot v_{(3)}) \, (n_{3}^{T} \cdot v_{(3)})$. At the 
order of approximation retained, 
$(e^{T} \cdot v_{(3)}) \sim {\cal O}(s_{-})$, 
$(n_{3}^{T} \cdot v_{(3)}) \sim {\cal O}(s_{-})$ and in
both terms the relevant contribution\footnote{Notice that 
$(n_{3}^{T} \cdot v_{(3) \, \emptyset}) = 
(e^{T} \cdot v_{(3) \, \emptyset}) = 0$.}  comes from the component 
$- \, s_{-}/(\widetilde{\lambda}\sqrt{2}) \; l_{(3)}$ of 
$v_{(3)}$ in eq. (\ref{eq14bis}). Since 
$( n_{3}^{T} \cdot l_{(3)} ) = 4 \, m^{2}$ we have:
\begin{eqnarray}
\sigma_{3}^{-1} \,
\left( e^{T} \cdot v_{(3)} \right) 
\left( n_{3}^{T} \cdot v_{(3)} \right)
& = &
\frac{4 \, m^{2}}{\widetilde{\lambda}} \, 
\frac{s_{-}^{2}}{\sigma_{3}} \;
\frac{1}{2 \, \widetilde{\lambda}} \, 
 \left( e^{T} \cdot l_{(3)} \right) + 
\cdots
\label{eq45}
\end{eqnarray}
The combination of eqs. (\ref{eq44b}) and (\ref{eq45}) involves:
\begin{eqnarray}
1 + \frac{4 \, m^{2}}{\widetilde{\lambda}} \, 
\frac{s_{-}^{2}}{\sigma_{3}} \;
& = &
- \, \frac{s_{2}}{\sigma_{3}} + \cdots
\label{eq46}
\end{eqnarray}
so that
\begin{eqnarray}
b_{3}(\sigma \to 0)
& = & 
- \, \frac{s_{2}}{\sigma_{3}} \,
\frac{1}{2 \, \widetilde{\lambda}} \, \left( e^{T} \cdot l_{(3)} \right)
+ \cdots
\label{eq47}
\end{eqnarray}

\vspace{0.3cm}

\noindent
{\bf (iii) $(b_{0})_{1} + (b_{0})_{2}$}

\vspace{0.3cm}

\noindent
We have:
\begin{eqnarray}
(b_{0})_{1}+(b_{0})_{2}
& = &
- \, \frac{\sigma_{2 \, \emptyset}}{2 \, \widetilde{\lambda}} \, 
\left( e^{T} \cdot v_{_{(1) \, \emptyset}} \right) 
\left( (n_{1} + n_{2})^{T} \cdot v_{_{(1) \, \emptyset}} \right) \, 
\nonumber\\
&& 
- \, \frac{\sigma_{1 \, \emptyset}}{2 \, \widetilde{\lambda}}  \, 
\left( e^{T} \cdot v_{_{(2) \, \emptyset}} \right) 
\left( (n_{1} + n_{2})^{T} \cdot v_{_{(2) \, \emptyset}} \right)
\label{eq48}
\end{eqnarray}
It involves
\begin{eqnarray}
\left( e^{T} \cdot v_{_{(1) \, \emptyset}} \right)  
\left( (n_{1} + n_{2})^{T} \cdot v_{_{(1) \, \emptyset}} \right)
& = &
\left( 
 \sqrt{2} \, \cos \theta_{\emptyset} \, -  \, \sin \theta_{\emptyset}
\right) \, \sqrt{2} \, \cos \theta_{\emptyset}
\label{eq49a}\\
\left( e^{T} \cdot v_{_{(2) \, \emptyset}} \right) 
\left( (n_{1} + n_{2})^{T} \cdot v_{_{(2) \, \emptyset}} \right)
& = &
 \left( 
 \sqrt{2} \, \sin \theta_{\emptyset} \, + \, \cos \theta_{\emptyset}
\right) \, \sqrt{2} \, \sin \theta_{\emptyset}
\label{eq49b}
\end{eqnarray}
$(b_{0})_{1}+(b_{0})_{2}$ takes the form:
\begin{eqnarray}
(b_{0})_{1}+(b_{0})_{2}
& = &
\! - \, \frac{1}{2 \, \widetilde{\lambda}} \, 
\Biggl\{
    \left( \sigma_{1 \, \emptyset} + \sigma_{2 \, \emptyset} \right)
\nonumber\\
&&
\;\;\;
  -
 \left[ 
  \left( \cos^{2} \theta_{\emptyset} - \sin^{2} \theta_{\emptyset} \right)
  - \sqrt{2} \, \sin \theta_{\emptyset} \, \cos \theta_{\emptyset}
 \right] \, \left( \sigma_{1 \, \emptyset} - \sigma_{2 \, \emptyset} \right)
\Biggr\}
\;\;\;\;\;\;
\label{eq50}
\end{eqnarray}
With $\cos \theta_{\emptyset}$, $\sin \theta_{\emptyset}$ from eqs. 
(\ref{eq17c}) - (\ref{eq18b}) and using identity (\ref{eq17suite}), we have:
\begin{eqnarray}
\lefteqn{
  \left( \cos^{2} \theta_{\emptyset} - \sin^{2} \theta_{\emptyset} \right)
  - \sqrt{2} \, \sin \theta_{\emptyset} \, \cos \theta_{\emptyset}
}
\nonumber\\
& = &
\frac{1}
{\left( \sigma_{1 \, \emptyset} - \sigma_{2 \, \emptyset} \right)} \, 
\left[
 2 \, \left(s_{+} - (m^{2} + m_{3}^{2}) \right) \, + \, 2 \, (2 \, m_{3}^{2}) 
 \, + \,
 \left( \sigma_{1 \, \emptyset} + \sigma_{2 \, \emptyset} \right)
\right]
\label{eq51b}
\end{eqnarray}
Finally $(b_{0})_{1}+(b_{0})_{2}$ reads:
\begin{eqnarray}
(b_{0})_{1}+(b_{0})_{2}
& = &
\frac{1}{\widetilde{\lambda}} \, 
\left[ 
 \left( s_{+} - (m^{2} + m_{3}^{2}) \right)
 \,+ \, (2 \, m_{3}^{2})
\right]
\label{eq52}
\end{eqnarray}
{\bf (iv) $(b_{1}+b_{2})(\sigma_{3} \to 0)$}

\vspace{0.3cm}

\noindent
The extra bit added to $(b_{0})_{1}+(b_{0})_{2}$ to obtain
$(b_{1}+b_{2})(\sigma_3 \to 0)$ is proportional to
$(e^{T} \cdot v_{_{(3)}}) \, ((n_{1}+n_{2})^{T} \cdot v_{_{(3)}})$. 
Here again\footnote{Notice that 
$( (n_{1}+n_{2})^{T} \cdot v_{_{(3)} \emptyset} ) = 0$.}, 
$(e^{T} \cdot v_{_{(3)}}) \sim {\cal O}(s_{-})$, 
$((n_{1}+n_{2})^{T} \cdot v_{(3)}) \sim {\cal O}(s_{-})$ and 
in both terms the relevant contribution comes from the component 
$- \,  s_{-}/(\widetilde{\lambda}\sqrt{2}) \; l_{(3)}$ of 
$v_{_{(3)}}$ in eq. (\ref{eq14bis}). The product of these two 
contributions provides the term sought. Since
$((n_{1}+n_{2})^{T} \cdot l_{(3)} ) = 2 \, (s_{+}-(m^{2}+m_{3}^{2}))$, 
we have:
\begin{equation}
\sigma_{3}^{-1}
\left( e^{T} \cdot v_{(3)} \right)
\left( (n_{1}+n_{2})^{T} \cdot v_{(3)} \right)
= \frac{1}{\sigma_{3}} \, 
\frac{s_{-}^{2}}{\widetilde{\lambda}}
\frac{ \left( s_{+}-(m^{2}+m_{3}^{2}) \right)}{\widetilde{\lambda}} 
 \left( e^{T} \cdot l_{(3)} \right) + \cdots \;\;\;
\label{eq53}
\end{equation}
Rewriting
\begin{equation}
\left( s_{+}-(m^{2}+m_{3}^{2}) \right) \, 
\left( e^{T} \cdot l_{(3)} \right)
= 
2 \, \widetilde{\lambda} \, + \, (4 \, m^{2})
\left[ \left( s_{+}-(m^{2}+m_{3}^{2}) \right) \, + (2 \, m_{3}^{2}) \right]
\label{eq54}
\end{equation}
we get:
\begin{eqnarray}
\lefteqn{
\sigma_{3}^{-1} \, \left( e^{T} \cdot v_{(3)} \right) 
\left( (n_{1}+n_{2})^{T} \cdot v_{(3)} \right)
}
\nonumber\\
& = &
\frac{2 \, s_{-}^{2}}{\sigma_{3} \, \widetilde{\lambda}} 
\, + \, 
\frac{s_{-}^{2}}{\sigma_{3}} \, \frac{4 \, m^{2}}{( \widetilde{\lambda})^2}
\left[ \left( s_{+}-(m^{2}+m_{3}^{2}) \right) \, + (2 \, m_{3}^{2}) \right]
+ \cdots
\label{eq55}
\end{eqnarray}
The combination of eqs. (\ref{eq52}) and (\ref{eq55}) using eq. 
(\ref{eq46}) leads to: 
\begin{eqnarray}
\lefteqn{(b_{1}+b_{2})(\sigma_{3} \to 0)}
\nonumber\\ 
& = & 
\frac{2 \, s_{-}^{2}}{\sigma_{3} \, \widetilde{\lambda}} 
- \, \frac{s_{2}}{\sigma_{3}} \,
\frac{1}{\widetilde{\lambda}} \, 
\left[ \left( s_{+}-(m^{2}+m_{3}^{2}) \right) \, + (2 \, m_{3}^{2}) \right]
+ \cdots
\label{eq56}
\end{eqnarray}

\vspace{0.3cm}

\noindent
{\bf (v) $B(\sigma_{3} \to 0)$}

\vspace{0.3cm}

\noindent
As a check, let us combine eqs. (\ref{eq47}) and (\ref{eq56}). We get:
\begin{eqnarray}
(b_{1}+b_{2}+b_{3})(\sigma_{3} \to 0)
& = &
\frac{2 \,  s_{-}^{2}}
{\widetilde{\lambda} \, \sigma_{3}} 
- \, 
\frac{s_{2}}{\widetilde{\lambda} \, \sigma_{3} } \, 
\Biggl\{
 \left[ \left( s_{+}-(m^{2}+m_{3}^{2}) \right) \, + (2 \, m^{2}) \right]
\nonumber\\
&& 
\;\;\;\;\;\;\;\;\;\;\;\;\;\;\;\;\;\;\;\;
+
 \left[ \left( s_{+}-(m^{2}+m_{3}^{2}) \right) \, + (2 \, m_{3}^{2}) \right]
\Biggr\} + \cdots
\nonumber\\
& = &
\frac{2^{2} \,  
\left( s_{+} s_{2}  - s_{-}^{2} \right) }
{- \, 2  \, \widetilde{\lambda} \, \sigma_{3}} 
+ \cdots
\label{eq57}
\end{eqnarray}
where we recognize the identity
\[
B = \frac{\detg}{\dets}
\]
for $\sigma_3 \to 0$ since the numerator and the denominator of the r.h.s. 
of eq. (\ref{eq57}) are the expressions of $\detg$ and $\dets$ 
respectively, at the approximation retained cf. eqs. (\ref{approxdetg}), 
(\ref{approxdetS}) and (\ref{eq5}).

\vspace{0.3cm}

\noindent
{\bf (viii) $B_{0}$}\vspace{0.3cm}

\noindent
Combining eqs. (\ref{eq44b}) and (\ref{eq52}) we also get
$B_{0} = (b_{0})_{1}+(b_{0})_{2}+(b_{0})_{3}$:
\begin{eqnarray}
B_{0}
& = & 
\frac{1}{\widetilde{\lambda}} \,
\left\{
 \left[ \left( s_{+} - (m^{2} + m_{3}^{2}) \right) + (2 \, m^{2}) \right]
 +
 \left[  \left( s_{+} - (m^{2} + m_{3}^{2}) \right) + (2 \, m_{3}^{2}) \right]
\right\} \;\;\;\;\;\;
\nonumber\\
& = &
\frac{2 \, s_{+}}{\widetilde{\lambda}}
\label{eq58}
\end{eqnarray}
Notice that $B_{0}$ happens to coincide with the limit $t \to \infty$ 
of $B$ seen as a function of $t = s_{2}/s_{-}^{2}$, as given by eq. 
(\ref{B-vs-t}) in Appendix \ref{appendix_b}. 

\vspace{0.3cm}

\noindent
{\bf (vii) $(b_{0})_{1} - (b_{0})_{2}$}

\vspace{0.3cm}

\noindent
Since $n_{1} - n_{2} = \sqrt{2} \, v_{(3) \, \emptyset}$, 
$\left( (n_{1} - n_{2})^{T} \cdot v_{(j) \, \emptyset} \right) = 0$, $j=1,2$ 
thus we have: 
\begin{eqnarray}
(b_{0})_{1}-(b_{0})_{2}
& = &
- \, \frac{\sigma_{2 \, \emptyset}}{2 \, \widetilde{\lambda}} \, 
\left( e^{T} \cdot v_{(1) \, \emptyset} \right) 
\left( (n_{1} - n_{2})^{T} \cdot v_{(1) \, \emptyset} \right) \, 
\nonumber\\
&& 
- \, \frac{\sigma_{1 \, \emptyset}}{2 \, \widetilde{\lambda}}  \, 
\left( e^{T} \cdot v_{(2) \, \emptyset} \right) 
\left( (n_{1} - n_{2})^{T} \cdot v_{(2) \, \emptyset} \right)
\nonumber\\
& = &
0
\label{eq59}
\end{eqnarray}
{\bf (viii) $(b_{1}-b_{2})(\sigma_{3} \to 0)$}

\vspace{0.3cm}

\noindent
Given eq. (\ref{eq59}), $(b_{1}-b_{2})(\sigma_{3} \to 0)$ is given by:
\begin{eqnarray}
(b_{1}-b_{2})(\sigma_{3} \to 0)
& = &
\sigma_{3}^{-1} \,
\left( e^{T} \cdot v_{(3)} \right) \, 
\left( (n_{1}-n_{2})^{T} \cdot v_{(3)} \right)
\label{eq60}
\end{eqnarray}
Whereas $(e^{T} \cdot v_{(3) \, \emptyset}) = 0$,
$((n_{1}-n_{2})^{T} \cdot v_{(3) \, \emptyset}) = \sqrt{2} \neq 0$. This makes 
$(b_{1}-b_{2})(\sigma_{3} \to 0)$ diverge. 
More precisely, since $((n_{1}-n_{2})^{T} \cdot l_{(3)}) = 0$, the 
${\cal O}(s_{-})$ terms in the r.h.s. of eq. (\ref{eq61}) cancel and, from
eq. (\ref{eq14bis}) and we get: 
\begin{eqnarray}
\left( (n_{1}-n_{2})^{T} \cdot v_{(3)} \right)
& = & 
\left( 1 + {\cal O}(s_{-}^{2}) \right) \, \sqrt{2} + \cdots
\label{eq61}
\end{eqnarray}
As $(e^{T} \cdot v_{(3)}) = {\cal O}(s_{-})$, the ${\cal O}(s_{-}^{2})$
correction in eq. (\ref{eq61}) leads to a contribution to 
$(b_{1}-b_{2})(\sigma_{3} \to 0)$ which is 
$\sim {\cal O}( s_{-}^{3}/ \sigma_{3})$ i.e. beyond the approximation retained. 
We thus keep:
\begin{eqnarray}
  (b_{1}-b_{2})(\sigma_{3} \to 0)
& = &
\sigma_{3}^{-1} \,
\left( e^{T} \cdot v_{(3)} \right) \, 
\left( (n_{1}-n_{2})^{T} \cdot v_{(3) \, \emptyset} \right) + \cdots 
\nonumber\\
& = & 
- \, \frac{s_{-}}{\widetilde{\lambda} \, \sigma_{3}} \, 
\left( e^{T} \cdot l_{(3)} \right) + \cdots 
\label{eq60}
\end{eqnarray}
with $( e^{T} \cdot l_{(3)} )$ given by eq. (\ref{eq44b}). This makes 
$(b_{1}-b_{2})(\sigma_{3} \to 0)$ diverge as $s_{-}/\sigma_{3}$ which is 
$\sim {\cal O}(s_{-}^{-1}) \sim {\cal O}(\sigma_{3}^{-1/2})$
when both $\detg$ and $\dets$ tend to zero.

%% file: appendix-zeromode.tex
\mysection{A relation between the zero eigenmodes of ${\cal S}$ 
and  $G^{(N)}$ when $\detg =0$, $\det ({\cal S}) =0$}\label{app-degen}

We specify $a = N$ to fix the ideas. When $\detg = 0$, 
condition (\ref{linearized}) is equivalent to the condition 
$\det ({\cal S})=0$ according to eq. (\ref{detsdetggtilde}) only if 
$G^{(N)}$ has rank $(N-2)$. On the other hand when $G^{(N)}$ has a lower rank, 
its cofactor matrix $\widetilde{G}^{(N)}$ vanishes identically thus 
$\det ({\cal S})=0$ again. However, as already mentioned,  Gram matrices 
$G^{(N)}$ with ranks $\leq (N-3)$ for $N=3,4$ 
correspond to quite peculiar and degenerate kinematics irrelevant 
for NLO processes, thus we do not provide any more detail 
about this case here. We focus on the generic case for which the Gram matrix 
has rank $(N-2)$ i.e. exactly one vanishing eigenvalue. 

\vspace{0.3cm}

\noindent
When $\detg$ and $\dets$ vanish simultaneously the eigenvectors
$v_{(N)}$ and $\hat{n}^{(N)}$ corresponding to the eigenvalues zero of ${\cal S}$ and
$G^{(N)}$ respectively, happen to be simply related. To see this,
using eqs. (\ref{eqGRAMMAT}), (\ref{eqVJA}) let us write the components $i=1,
\cdots, N$  of ${\cal S} \cdot v$ for any $N$-column vector $v$ as:
\begin{eqnarray}
\left( {\cal S} \cdot v \right)_{i} 
\!\!
& = &
\!\!
\left\{
\!\!
\mbox{\small
$ \begin{array}{cl}
  \left( \sum_{j=1}^{N-1} G^{(N)}_{ij} \, v_{j} \right) + 
  \left( \sum_{j=1}^{N-1} V^{(N)}_{j} \, v_{j} \right) + 
  \left( V^{(N)}_{i} + S_{NN} \right) 
  \left( \sum_{j=1}^{N} v_{j} \right) & \mbox{if} \; i \neq N\\
  \left( \sum_{j=1}^{N-1}V^{(N)}_{j} \, v_{j} \right) + 
  S_{NN} \left( \sum_{j=1}^{N} v_{j} \right) & \mbox{if} \; i = N 
 \end{array}
$}
\right.
\nonumber\\
&&
\label{e9cons}
\end{eqnarray}
As argued in subsubsec. \ref{sss221}, the eigenvector $\hat{n}^{(N)}$ fulfills 
condition (\ref{linearized}). Therefore the ansatz
\begin{eqnarray}
v_{(N) \, j} \equiv \hat{n}^{(N)}_{j}, \;\; j = 1, \cdots, N-1
& , &
v_{(N) \, N} \equiv  - \, \sum_{j=1}^{N-1} \hat{n}^{(N)}_{j} 
\end{eqnarray}
is solution of system (\ref{e9cons}). Furthermore it satisfies the 
property $\sum_{j=1}^{N} v_{(N) \, j} = 0$ by construction. If ${\cal S}$ has 
rank $(N-1)$, 
the eigendirection of ${\cal S}$ associated to the eigenvalue zero is thus 
identified.

\vspace{0.3cm}

\noindent
Conversely, consider $v$ such that 
\begin{eqnarray}
\sum_{j=1}^{N} {\cal S}_{ij} \, v_{j} & = & 0
\label{converse}
\end{eqnarray}
and define 
\begin{eqnarray}
\delta & \equiv &  \sum_{j=1}^{N} v_{j} 
\label{norm}
\end{eqnarray}
Using eq. (\ref{e9cons}), eqs. (\ref{converse}), (\ref{norm}) may be written:
\begin{eqnarray}
\sum_{j=1}^{N-1} G^{(N)}_{ij}\, v_{j} & = & \delta \, V^{(N)}_{i},  
\;\; i=1,\cdots, N-1 
\label{e7bis-a}\\
\sum_{j=1}^{N-1} V^{(N)}_{j} \, v_{j} & = & - \, \delta \, S_{NN}
\label{e7bis-b}
\end{eqnarray}
If $\delta = 0$, the $(N-1)$-column vector 
$\hat{n}_{i} \equiv v_{i}$, $i = 1, \cdots, N-1$ is an eigenvector of $G^{(N)}$ 
associated to the eigenvalue zero and fulfilling condition (\ref{linearized}).

\vspace{0.3cm}

\noindent
Let us conclude these considerations with the following remarks. 
\begin{enumerate}
\item
We recall that, at the one loop order which we are concerned with, the 
Landau conditions \cite{kinem,landaucond} characterizing the appearance of kinematic 
singularities require $v_{i} \geq 0$ for all $i = 1, \cdots, N$ and 
$\delta > 0$. One may hastily infer that, an eigendirection zero of 
${\cal S}$ associated with a vanishing $\detg$ is not associated with a 
kinematic singularity as characterized by the Landau conditions.
This does hold true if ${\cal S}$ has rank $(N-1)$. 
\item
Let us however note that if ${\cal S}$ has two vanishing eigenvalues with
corresponding linearly independent eigenvectors $v_{(1)}$ and $v_{(2)}$ both 
such that $\sum_{j=1}^{N} v_{(i) \, j} \neq 0$, their components can be 
rescaled in order that 
$\sum_{j=1}^{N} v_{(1) \, j} = \sum_{j=1}^{N} v_{(2) \, j} = \delta$.
The $(N-1)$-column vector $\hat{n}$ defined by
$\hat{n}_{i} =v_{(1) \, i} \, - \, v_{(2) \, i}, i=1, \cdots, N-1$ then
fulfills\footnote{If ${\cal S}$ has rank $\leq N-2$, such eigenvectors 
$v_{(1)}$ and $v_{(2)}$ can always be found even if ${\cal S}$ is complex. 
In the latter case, consider linearly independent eigenvectors $u_{1}$ and 
$u_{2}$ of $\cals \, \cals^{\dagger}$ associated with the eigenvalue zero: their 
respective complex conjugates $v_{(1)}=u_{1}^{*}$ and $v_{(2)}=u_{2}^{*}$ 
are linearly independent eigenvectors of ${\cal S}$ associated with the 
eigenvalue zero. The matrix 
$G^{(N)}$ being symmetric real, the eigenvector $\hat{n}$ of $G^{(N)}$ 
built as described shall be made real by an overall phase shift.} 
$G^{(N)} \cdot \hat{n} = 0$ and condition (\ref{linearized}). 

\vspace{0.3cm}

\noindent
In particular, $v_{(1)}$ and $v_{(2)}$ may both fulfill the Landau conditions
corresponding to piled-up singularities. The so-called double parton 
singularity \cite{dps} is one interesting case of this kind. It occurs for the 
four-point function with opposite lightlike and opposite timelike legs and 
with internal masses all vanishing, for which 
$\det \, ({\cal S}) \propto \detg^2$, 
when the two lightlike momenta are incoming 
head-on and the two timelike external momenta are outgoing back-to-back 
in the transverse plane w.r.t. the incoming direction\footnote{The fact that 
such configuration leads to a vanishing $\detg$ does not contradict the 
classification of the eligible kinematics provided in appendix 
\ref{appendix-a2}. This appendix focused on the kinematical configurations
corresponding to linearly {\em independent} sets of four-momenta. On the 
contrary the double parton scattering singularity appears for coplanar 
configurations of linearly {\em dependent} four-momenta.}.

\item
In practice we shall however stress that such a degeneracy of the zero 
eigenvalue of ${\cal S}$ 
is very peculiar. Beside the double parton singularity, this situation happens 
to occur for $N=4$ with three of the four internal 
masses equal, for very peculiar degenerate kinematics involving two 
spacelike momenta, and two lightlike momenta orthogonal to the spacelike ones, 
collinear to each other and being linear combination of the spacelike ones... 
This quite degenerate kinematics namely corresponds to a Gram matrix with rank 
1 only. As already said, such an odd case is actually irrelevant regarding NLO 
processes at colliders, thus do not deserve any further detail here.  
\end{enumerate}

%% file: vep_appendix_b.tex
\mysection{The directional limit $\dets \to 0$, 
$\detg$ of eq. (\ref{recombI3}) is actually isotropic}\label{appendix_b}

This appendix presents an analytical proof that, whereas each of the three 
terms involved in  eq. (\ref{recombI3}) are separately functions  of $t$ in 
the directional limit $s_{-} \to 0$, $s_{2} \to 0$ with 
$s_{2}  / s_{-}^{2} = t$ fixed, the limit of their {\em sum} is 
actually {\em independent} of $t$. For this purpose we compute the 
$t$-derivative of this sum in this limit and prove it to vanish identically in 
$t$. We provide an explicit proof for $I_{3}^{4}$; the $I_{3}^{6}$ case, albeit 
more cumbersome, can be handled in a completely similar way.

\vspace{0.3cm}

\noindent
In the limit $s_{-} \to 0$, $s_{2} \to 0$, 
$s_{2}  / s_{-}^{2} = t$ fixed, eq. (\ref{recombI3}) reads:
\begin{eqnarray}
\lefteqn{\sum_{j=1}^{3} b_{j} \, {\cal I}_{3 \, (j)} }
\nonumber\\
& = & 
- \Biggl[ 
\frac{t \, (s_{+} + \, (m^2-m_3^2))}
{(t \, \widetilde{\lambda} + 4 \, m^2)} \,
\int^1_0 dz \, \frac{\ln(m^2)-\ln(S_0)}{2 \, B \, m^2 + 1} 
\nonumber \\
& & 
\mbox{} +
\frac{t \, (s_{+} +\, (m_3^2-m^2)) - 2}{(t \, \widetilde{\lambda} + 4 \, m^2)} 
\, \int^1_0 dz \,  \frac{\ln(g(z))-\ln(S_0)}{2 \, B \, g(z) + 1} 
\nonumber \\
& & 
\mbox{} + 
\frac{- \, s_{+} + \, (m_3^2-m^2)}{(t \, \widetilde{\lambda} + 4 \, m^2)} \, \int^1_0 dz \, 
\nonumber\\
& & \mbox{} \times
\Biggl\{ 4 \, B \, z \, (1-z) \, 
\frac{\ln(g(z)) - \ln(S_0)}{(2 \, B \, g(z) + 1)^2} 
- \frac{z \, (1-z)}{g(z)} \, \frac{2}{2 \, B \, g(z) + 1} 
\Biggr\} 
\Biggr]
\label{startdifft}
\end{eqnarray}
where
\begin{eqnarray}
B 
& = & 
- \, \frac{2 \, (1 - t \, s_{+})}{t \, \widetilde{\lambda} + 4 \, m^2},
\;\;\;\; 
S_{0} = - \frac{1}{2 \, B} - i \, \lambda 
\label{B-vs-t}\\
\widetilde{\lambda} 
& = & 
\left( s_{+} - (m^2+m_3^2) \right)^2 - 4 \, m^2 \, m_3^2
\label{lambda=alternative}\\
g(z) 
& = & 
s_{+} \, z^2 + 
\left( - s_{+} + m^2 - m_3^2 \right) \, z + m_3^2
\label{altdefg}
\end{eqnarray}
The kinematic parameter $\widetilde{\lambda}$ was already defined by eq. 
(\ref{lambda}), the equivalent form (\ref{lambda=alternative}) is given here 
for convenience, as eq. (\ref{altdefg}) does for the function $g(z)$ 
previously defined by eq. (\ref{limg}) and fulfilling eqs. (\ref{limg1}) 
and (\ref{limg2}).
To keep formulas compact, let us introduce the following notations: 
$H(z,t) = \ln(g(z))-\ln(S_0)$, 
$D(z,t) = - 4 \, (1 - t \, s_{+}) \, g(z) + t \, \widetilde{\lambda} + 4 \, m^2$, 
$\Delta_m = (m_3^2 - m^2)$, $T_1 = (1 - t \, s_{+})$, 
$T_2= \Delta_m - s_{+}, T_3= \Delta_m + s_{+}$.
Differentiating eq. (\ref{startdifft}) w.r.t. $t$ leads to:
\begin{equation}
  \frac{d}{d t} \,\sum_{j=1}^{3} b_j \, {\cal I}_{3 \, (j)} = - \left[ P_1 + P_2 +P_3 +P_4 + P_5 \right]
\label{eqAB1}
\end{equation}
with
\begin{eqnarray*}
P_1 
& = & 
\frac{1}{4} \, \frac{T_2}{T_1^2 \, S_0 } 
\\
P_2 
& = & 
\left( 1+T_1- t \, \Delta_m \right) \, 
\int_0^1 dz \, \frac{(g'(z))^2 \, H(z,t)}{D(z,t)^2} 
\\
P_3 
& = & 
\frac{1}{4 \, S_0 \,T_1^2} \, 
\int_0^1 dz \, 
\frac{T_2^2 \, (1 + T_1 -  t \, \Delta_m) + 4 \, H(z,t) \, T_3 S_0 \, T_1^2}
{D(z,t)} 
\\
P_4 
& = & 
\frac{2 \, T_2}{T_1 \, S_0 } \, 
\int_0^1 dz \, 
\frac{z \, (1-z) \,  (g(z) \, T_2^2 + 4 \, s_{+} \, g(z) \, S_0 \, T_1 \, H(z,t) + 
(g'(z))^2 \, S_0 T_1)}
{g(z) \, D(z,t)^2}  
\\
P_5 
& = & \, 
16 \, T_1 \, T_2 \, 
\int_0^1 dz \, \frac{z \, (1-z) \, (g'(z))^2 \, H(z,t)}{D(z,t)^3}
\end{eqnarray*}
where $g'(z) = d g(z)/d z$.
To derive eq. (\ref{eqAB1}), we have used that:
\[
  \frac{\partial D(z,t)}{\partial t} =  (g'(z))^2
\]
We will not compute any of these integrals over $z$ explicitly: we will 
instead integrate by parts to iteratively decrease the powers of 
$D(z,t)$ in denominators, starting with $P_{5}$ which involves the highest 
power, and proceed to a step by step cancellation of terms on the way. 
For this purpose we note that the partial $z$-derivative 
of $D(z,t)$ is $g'(z)$ times a $z$-independent factor:
\begin{equation}
\frac{\partial D(z,t)}{\partial z} = - 4 \, T_1 \, g'(z)
\label{eqAB6}
\end{equation}
Integrating $P_{5}$ by parts and noticing that the boundary term vanishes due 
to the $z \, (1-z)$ factor, we get :
\begin{eqnarray}
P_5 
& = & 
2 \, T_2 \, 
\int_0^1 dz \, 
\Bigl\{ 
 H(z,t) \, g'(z) \, (2 \, z -1) - 2 \, s_{+} \,  z \, (1-z) \, H(z,t)
\nonumber \\
& & 
\;\;\;\;\;\;\;\;\;\;\;\;\;\;\;\;\;\;
- \, \frac{z \, (1-z) \, (g'(z))^2}{g(z)} 
\Bigr\} \, \frac{1}{D(z,t)^2}
\label{step1}
\end{eqnarray}
Accounting for eq. (\ref{step1}), let us now collect all the terms with 
denominator $D(z,t)^{2}$ in eq. (\ref{eqAB1}). We get:
\begin{eqnarray}
P_2 + P_4 + P_5 & = & \int_0^1 dz \, 
\frac{2 \, z \, (1-z) \, T_2^3 - S_0 \, T_1 \, T_3 \, D(z,t) \, H(z,t)}
{T_1 \, S_0 \, D(z,t)^2}
\label{eqAB2}
\end{eqnarray}
Comparing eq. (\ref{eqAB2}) and the equation which gives $P_3$ above, we see 
that the contribution proportional to $H(z,t)$ cancels out in the sum 
$\sum_{i=2}^{5} P_i$ which reads:
\begin{equation}
\sum_{i=2}^{5} P_i 
= 
\frac{T_2^2}{4 \, S_0 \, T_1^2} \, 
\int_0^1 dz \, 
\frac{8 \, T_1 \, T_2 \, z \, (1-z) + D(z,t) \, (1+T_1- t \, \Delta_m)}
{D(z,t)^2}
\label{eqAB3}
\end{equation}
To further decrease the power of $D(z,t)^2$ in eq. (\ref{eqAB3}), we notice 
that
\begin{eqnarray}
z \, (1-z)
& = & 
- \, \frac{1}{4 \, s_{+}^{2} \, T_2} \, 
\Bigl( 
\left( 2 \, \Delta_m - 
 \Delta_m \, t \, s_{+} - t \, s_{+}^{2} \right) \, (g'(z))^2 
\nonumber\\
& &
\;\;\;\;\;\;\;\;\;\;\;\;\;\;\;\;\;\;\;\;\; \mbox{}
 + 2 \, \Delta_m \, T_2 \, g'(z) + T_3 \, s_{+} \, D(z,t)
\Bigr)
\label{eqAB4}
\end{eqnarray}
Inserting eq. (\ref{eqAB4}) in eq. (\ref{eqAB3}), we get:
\begin{eqnarray}
\sum_{i=2}^{5} P_i & = & Q_1 + Q_2 + Q_3
\label{stepx}
\end{eqnarray}
with
\begin{eqnarray*}
Q_1 
& = & 
\frac{T_2^2 \, 
\left( s_{+}^{2} \, t + \Delta_m \, t \, s_{+} - \, 2 \, 
\Delta_m \right)}{2 \, T_1 \, S_0 \, s_{+}^{2}} \, 
\int_0^1 dz \, \frac{(g'(z))^2}{D(z,t)^2} 
\\ 
Q_2 
& = & 
- \, \frac{T_2^3 \, \Delta_m}{T_1 \, S_0 \, s_{+}^{2}} \, 
\int_0^1 dz \, \frac{g'(z)}{D(z,t)^2} 
\\
Q_3 
& = & 
- \, \frac{T_2^2}{4 \, T_1^2 \, S_0 \, s_{+}} \, 
\int_0^1 dz \, 
\frac{2 \, T_1 \, T_3 - s_{+} \, ( 1 + T_1 - \Delta_m \, t) }{D(z,t)}
\end{eqnarray*}
Again, an integration by parts of $Q_1$ and $Q_2$ using eq. (\ref{eqAB6}) gives:
\begin{eqnarray*}
Q_1 
& = & 
\frac{T_2^2 \, (s_{+}^{2} \, t + \Delta_m \, s_{+} \, t - 2 \, \Delta_m)}
{8 \, T_1^2 \, S_0 \, s_{+}^{2}} \, 
\Bigl( 
 \frac{g'(1)}{D(1,t)} - \frac{g'(0)}{D(0,t)} - 
 \int_0^1 dz \, \frac{2 \, s_{+}}{D(z,t)} 
\Bigr) 
\\ 
Q_2 
& = & 
- \, \frac{T_2^3 \, \Delta_m}{4 \, T_1^2 \, S_0 \, s_{+}^{2}} \, 
\Bigl( \frac{1}{D(1,t)} - \frac{1}{D(0,t)} \Bigr)
\end{eqnarray*}
The integrals of terms proportional to $1/D(z,t)$ in $Q_1$ and $Q_3$ cancel 
against each other. Besides, the definitions of $D(z,t)$ and $g'(z)$ lead to
\begin{eqnarray}
g'(1)= -\, T_2
& , &
g'(0) = -\, T_3
\\
D(1,t) = t \, T_2^2
& , &
D(0,t) = t \, T_3^2 - 4 \, \Delta_m
\end{eqnarray}
Substituting in eq. (\ref{stepx}), we find:
\begin{eqnarray}
\sum_{i=2}^{5} P_i & = & - \, \frac{T_2}{4 \, T_1^2 \, S_0} \; = \; - \, P_{1}
\end{eqnarray}
Hence
\begin{equation}
\frac{d}{d t} \,\sum_{j=1}^{3} b_{j} \, {\cal I}_{3 \, (j)} = 0
\label{eqAB5}
\end{equation}
q.e.d.